\documentclass{aa}  

\usepackage{natbib}
\usepackage[breaklinks=true, colorlinks=true, linkcolor=blue, citecolor=blue]{hyperref}
\bibpunct{(}{)}{;}{a}{}{,}

\usepackage{subfig}
\usepackage{graphicx}
\usepackage{adjustbox}
\usepackage{array,multirow,graphicx}
\usepackage{float}
\usepackage[dvipsnames]{xcolor}
\usepackage[normalem]{ulem}
\usepackage{lscape}
\usepackage{txfonts}
\usepackage{tablefootnote}

\makeatletter
\renewcommand*\aa@pageof{, page \thepage{} of \pageref*{LastPage}}
\makeatother
\defcitealias{Xu2018}{X18}
\defcitealias{Xu2022}{X22}

\begin{document} 

   \title{XMM-Newton follow-up of a sample of apparent low surface brightness galaxy groups detected in the ROSAT All-Sky Survey}
    \titlerunning{XMM-Newton follow-up of galaxy groups}
    \authorrunning{Spinelli et al.}
   \author{C. Spinelli, \inst{1}
          A. Veronica, \inst{1}
          F. Pacaud, \inst{1}
          T.H. Reiprich, \inst{1}
          K. Migkas, \inst{1,5}
          W. Xu \inst{2,3}
          \and
          M. E. Ramos-Ceja \inst{4}
          }
  \institute{Argelander-Institut f\"ur Astronomie (AIfA), Universit\"at Bonn, Auf dem H\"ugel 71, 53121 Bonn, Germany\\
              \email{averonica@astro.uni-bonn.de}
        \and
National Astronomical Observatories (NAOC), Chinese Academy of Sciences, Beijing 100101, China
  \and
School of Physics and Astronomy, Beijing Normal University, Beijing 100875, China
        \and
        Max-Planck-Institut f\"ur extraterrestrische Physik, Gießenbachstraße 1, 85748 Garching, Germany
\and
Leiden Observatory, Leiden University, PO Box 9513, 2300 RA Leiden, The Netherlands
        }

   \date{Received 1 March 2023 / Accepted 18 June 2025}

  \abstract
   {Galaxy cluster cosmology relies on complete and pure samples spanning over a large range of masses and redshifts. In \citet{Xu2018} and \citet{Xu2022},
   we discovered an apparently new population of galaxy groups and clusters with, on average, flatter X-ray surface brightness profiles than known clusters; this cluster population was missed in previous cluster surveys.
   The discovery of such a new class of objects could have a significant impact on cosmological applications of galaxy clusters.}
   {In this work we aim to characterize a subsample of these systems to assess whether they belong to a new population.}
   {We follow up three of these galaxy groups and clusters with high-quality XMM-Newton observations. We produce clean images and spectra and use them for model fitting. We also identify known galaxies, groups, or clusters in the field.}
   {The observations reveal that all three systems are composed of multiple groups each, either at the same or at different redshifts. In total, we characterize nine groups. We measure flat surface brightness profiles with slope parameter $\beta<0.6$; i.e, less than the canonical $\beta=2/3$. For the two main central groups, we even measure $\beta<0.4$.
   When the fluxes for the three observations are split up across the nine identified groups, none of them exceeds the typical flux limit adopted in previous RASS cluster catalogs, $\approx 3 \times 10^{-12}$ erg s$^{-1}$ cm$^{-2}$ in the 0.1--2.4 keV energy band.}
   {The observations reveal that groups with flat surface brightness profiles exist. Whether they form a new, separate population requires additional follow-up observations of further systems from the \citeauthor{Xu2022} sample, given the complexity we have discovered. Such extended low surface brightness systems, as well as multiple systems and projection effects, need to be taken into account when determining selection functions of group and cluster samples.}


   \maketitle
\section{Introduction}

   Galaxy groups and clusters are the largest virialized objects in the Universe. They represent a powerful probe of cosmological models \citep[e.g.,][]{2017MNRAS.471.1370S,2024PhRvD.110h3510B,2024A&A...689A.298G}, independent of other approaches such as those that make use of Type Ia Supernovae \citep[e.g.][]{SNIa2008} and of the cosmic microwave background \citep[CMB; e.g.][]{CMB}.
   
   Cluster masses can be inferred with several independent methods  \citep[e.g.,][]{SchellenbergerReiprich,Pratt2019,2024A&A...687A.178G,2024arXiv240208456K,2025arXiv250309952O} and they can be used to build the Halo Mass Function \citep[HMF; e.g.][]{Press&Schechter, Sheth&Tormen, Tinker_2008}. The HMF can be associated with the number density of galaxy clusters with a given mass, and its shape, normalization, and redshift evolution depend strongly on cosmology.
   In order to use galaxy groups and clusters for cosmological applications it is crucial to have well-defined, large samples of clusters covering a wide range of masses and redshifts that are representative of the underlying population \citep[e.g.,][]{Reiprich2002, Allen_2007, Vikhlinin2009, PlanckSZ,2024A&A...685A.106B}. Clearly, accurate knowledge of completeness and purity is key as the fundamental observable is the cluster number density \citep[e.g.,][]{2022A&A...665A..78S,2024A&A...687A.238C}.

   Some weak tensions have existed for a long time between CMB measurements and some galaxy cluster measurements of the cosmological parameters $\Omega_{\rm M}$ and $\sigma_8$ (e.g., \citealt{2006A&A...453L..39R} and references therein and Fig.~17 in \citealt{Pratt2019}). Usually, these tensions are referred to as $S_8$ tension\footnote{$S_8 = \sigma_8 \times \sqrt{\frac{\Omega_{\rm M}}{0.3}}$}, namely too few galaxy groups and clusters are detected for the given best-fit CMB values of $\Omega_{\rm M}$ and $\sigma_8$; i.e., the mean normalized matter density and its fluctuation amplitude on a linear scale of spheres with $8\,h^{-1}\,\mathrm{Mpc}$ radius, where $h$ is the normalized Hubble parameter. This raises the question whether clusters or especially groups might have  been missed by surveys systematically.
   
   Galaxy clusters can be observed in a number of wavebands, however the X-ray band allows us to observe and study the most massive visible component of clusters, the intracluster medium (ICM). The ICM is made up of gas reaching temperatures up to about $T_{  \text{gas}} = 10^8 $K which emits mostly through thermal bremsstrahlung in the X-ray band, proportional to density squared. The detection of this gas allows us to trace the presence of gravitationally bound structures and makes the X-ray imaging one of the most reliable methods to detect clusters and groups because it is least effected by projection effects.  

   In \citet{Xu2018} and \citet{Xu2022} (hereafter \citetalias{Xu2018} and \citetalias{Xu2022} respectively) we set out to test if 
   cluster surveys based on the ROSAT All-Sky Survey (RASS), a survey carried out with the ROSAT telescope with spatial resolution $\sim$$1'$ and low particle background, might have failed to detect a class of galaxy groups characterized by an unusually large extension, low surface brightness, or flat surface brightness profile. These characteristics would make it hard for such groups to be revealed with the detection method commonly used for most of the cluster samples built from RASS data. In fact, the sliding cell algorithm \citep{Harnden1984} is well suited to detect point sources but it may fail in identifying extended, low surface brightness objects. By applying to RASS observations the detection method based on wavelet-filtering presented in \cite{Pacaud2006}, many low redshift galaxy groups were identified at positions where no X-ray source was previously detected. A pilot sample of 13 such objects was presented in \citetalias{Xu2018}.
   The full sample includes 944 clusters among which 303 were discovered through their ICM emission for the first time and 149 of them had never been detected as galaxy clusters before in any waveband and is described in \citetalias{Xu2022}.
   If confirmed, the incompleteness of group and cluster samples could be higher than estimated, which could have important consequences for cosmology, as discussed in detail in \citet[their Section 4.3.1]{2017MNRAS.471.1370S}. Therefore, we study here a subsample of these systems with high-quality XMM-Newton follow-up observations.
   
\begin{figure*}[p!]
\centering
\vspace{-0.5cm}
\subfloat[RXGCC~127 -- RASS image]{\includegraphics[viewport=45 93 640 690, clip,height=6.5cm]{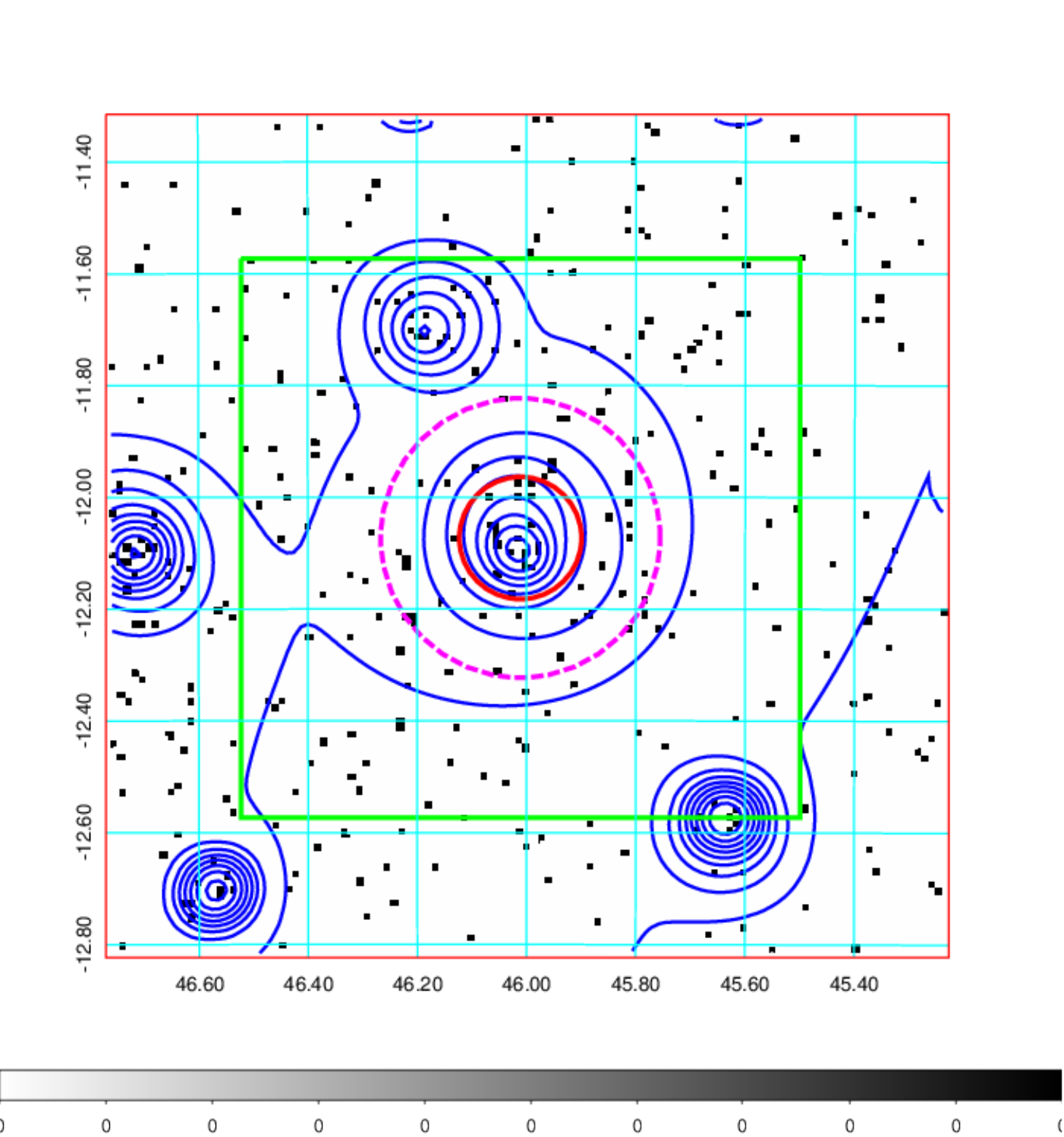}\label{fig:ROSAT127}}\hskip5ex
\subfloat[RXGCC~127 -- XMM-Newton image]{\includegraphics[height=6.5cm]{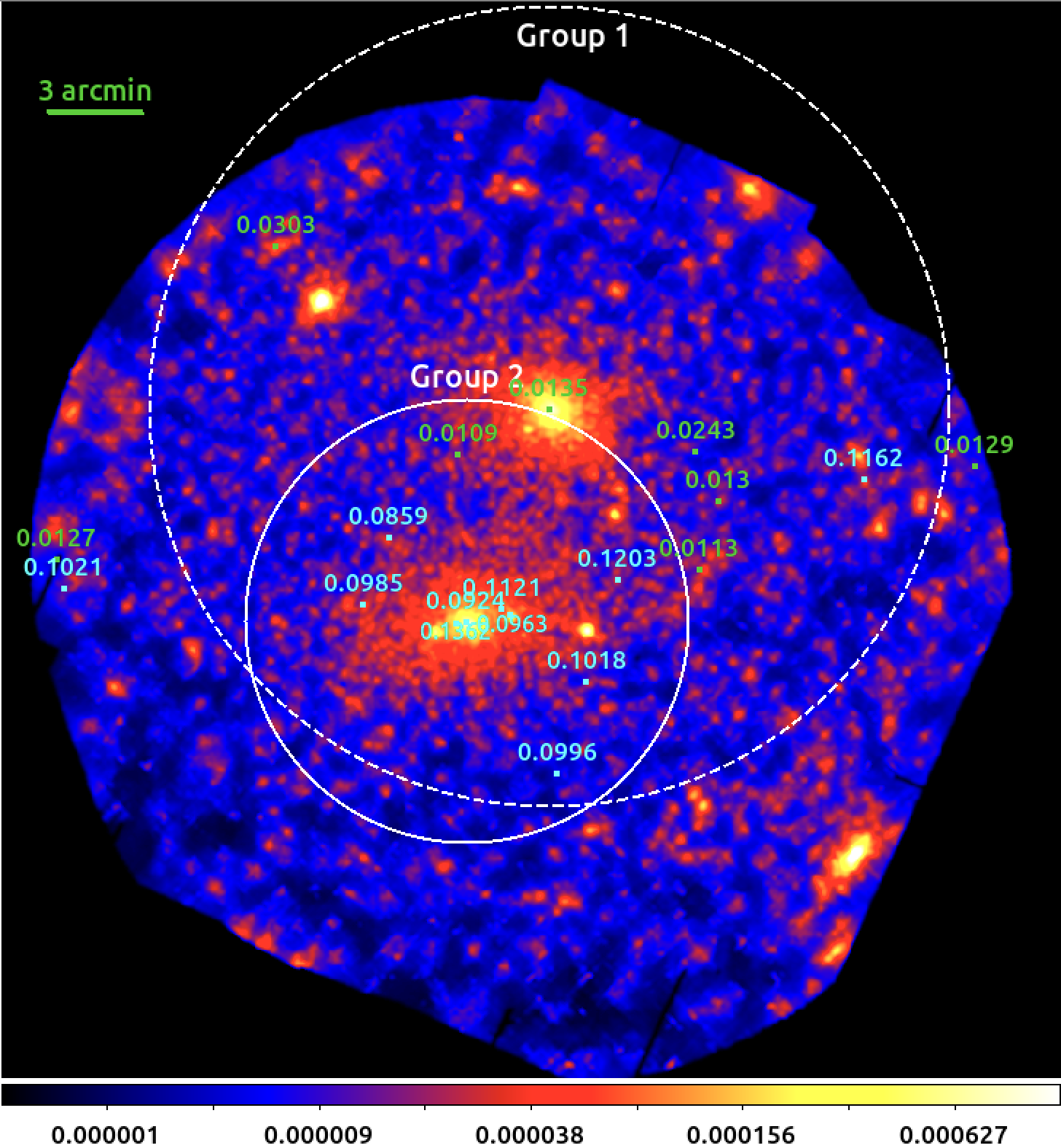}\label{fig:XMM127}}
\vspace{-0.3cm}\\
\subfloat[RXGCC~841 -- RASS image]{\includegraphics[viewport=45 93 640 690, clip,height=6.5cm]{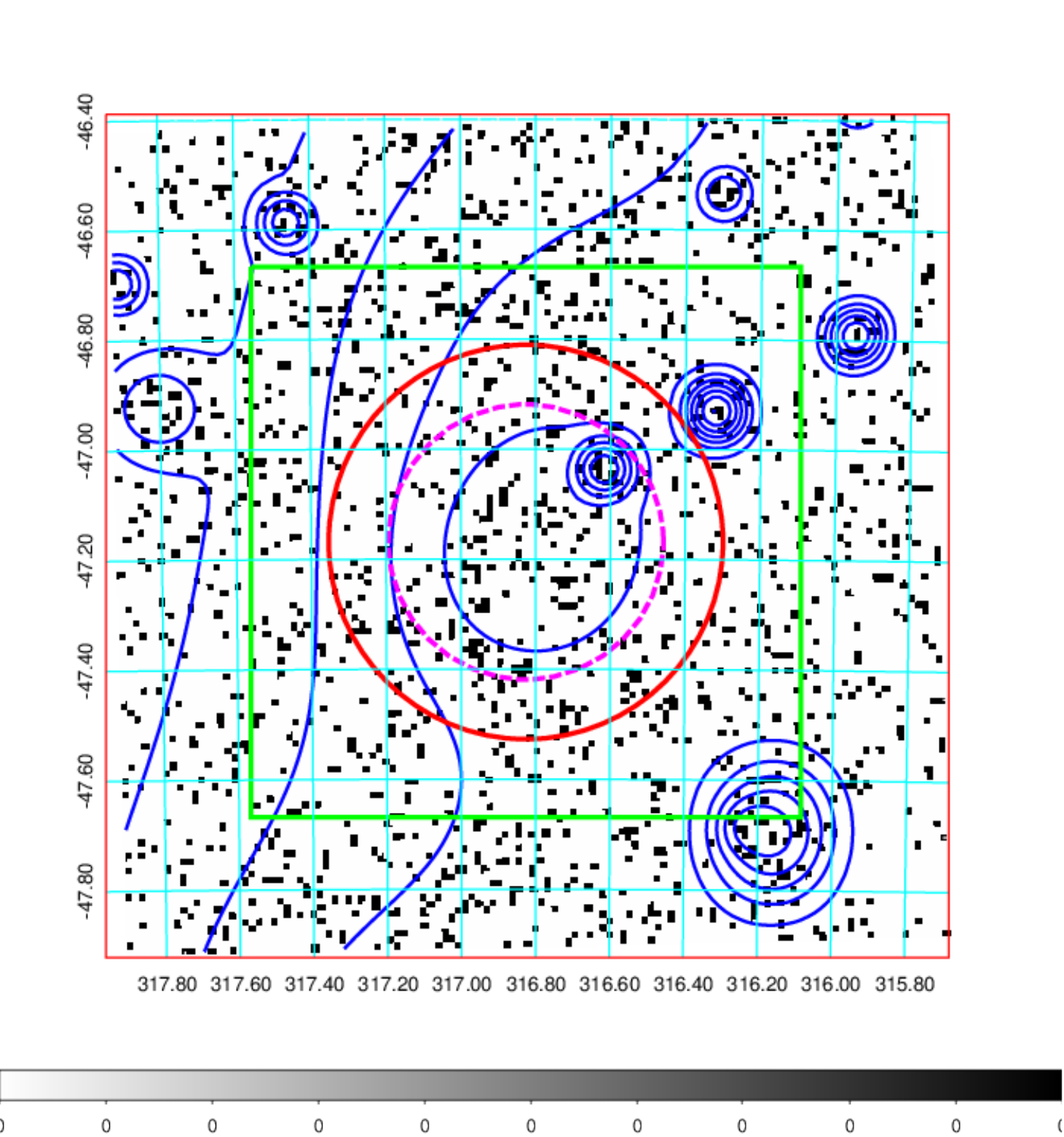}\label{fig:ROSAT841}}\hskip5ex
\subfloat[RXGCC~841 -- XMM-Newton image]{\includegraphics[height=6.5cm]{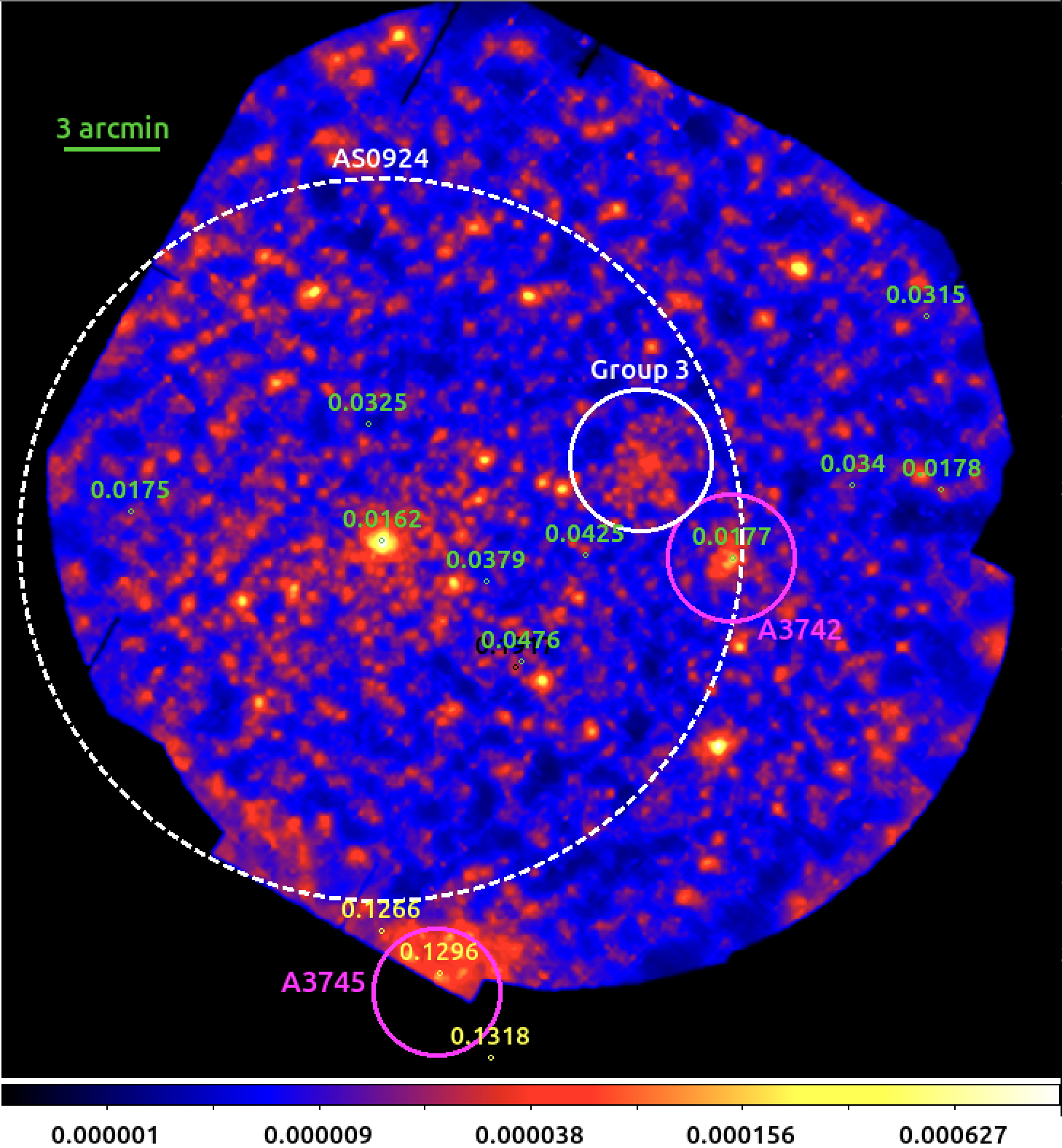}\label{fig:XMM841}}
\vspace{-0.3cm}\\
\subfloat[RXGCC~104 -- RASS image]{\includegraphics[viewport=45 93 640 690, clip,height=6.5cm]{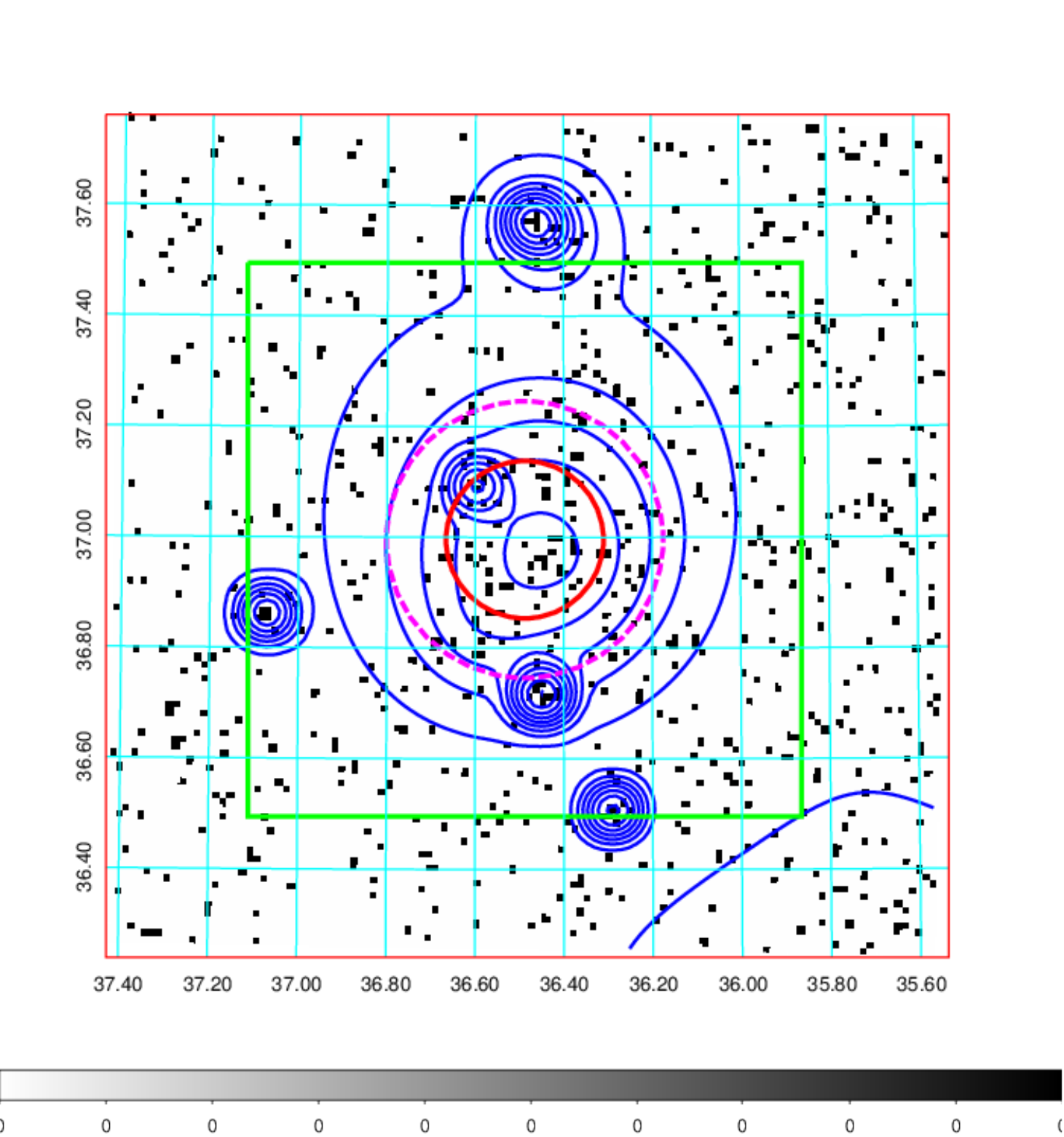}\label{fig:ROSAT104}}\hskip5ex
\subfloat[RXGCC~104 -- XMM-Newton image]{\includegraphics[height=6.5cm]{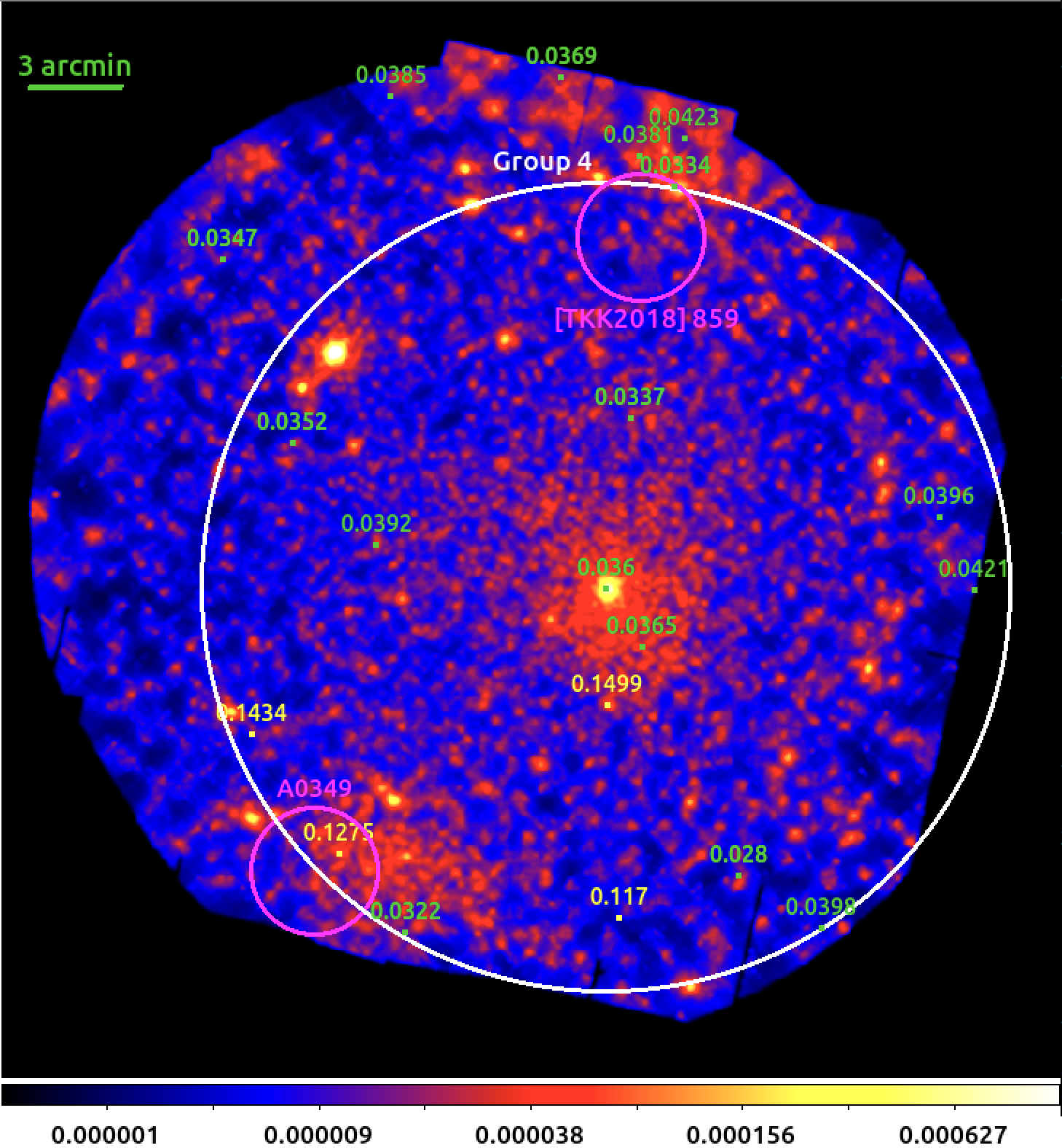}\label{fig:XMM104}}
\vspace{-0.1cm}
\caption{The red circle in the RASS photon images (left, in the 0.5--2 keV energy band) indicates the position and extent of the detection from \citetalias{Xu2022}. 
The magenta dashed circle shows the field-of-view of the XMM-Newton observation, and the green box serves as a $1^{\circ}{\times}1^{\circ}$ ruler. The blue contours show the smoothed signal from the wavelet filtered images used to detect the source. The XMM-Newton count-rate images (right, 0.4--1.25 keV) are instrumental background subtracted and smoothed by a 7.5$^{\prime\prime}$ Gaussian kernel. The labels indicate the position of the main sources discussed in the paper. The white circles mark the $R_{500}$ ($0.5\,R_{500}$ when dashed) of the groups that we determined in this paper (Sect.~\ref{sec:SpectralAnalysis}). Magenta circles around known galaxy clusters have an arbitrary size of $2'$. Matching known galaxies from NED are marked with small circles of various colors and used for group and cluster redshift estimates (Sect.~\ref{sect:imaging}): Panel~(b): for Group 1 in green and for Group 2 in cyan. Panel~(d): for Group 3 none, for AS0924 and A3742 in green, and for A3745 in yellow.  Panel~(f): for Group 4 and [TKK2018] 859 in green, and for A0349 in yellow.}
\label{fig:AllImages}
\end{figure*}

   We selected four of the galaxy groups from the pilot sample, whose X-ray fluxes are close to the nominal flux-limits of previous RASS cluster catalogs ($\approx 3 \times 10^{-12}$ erg s$^{-1}$ cm$^{-2}$ in the $0.1$--$2.4$ keV energy band), to be followed up by XMM-Newton. The observation of one them (RXGCC 507/Observation 4) turned out to be completely flared, therefore, we focus here on RXGCC 127, RXGCC 841 and RXGCC 104 in \citetalias{Xu2022}, which in this work will be referred to as Observation 1, 2, and 3 respectively. Their observation IDs, coordinates and estimated redshifts according to the RASS detection are summarized in Table~\ref{table:RaDec}. Fig.~\ref{fig:AllImages} shows the RASS and XMM-Newton images for our targets plus additional information as described in the caption.
   
   \begin{table}
\caption{XMM-Newton Observation ID, clean exposure time, coordinates,  and estimated redshift for our sample according to the RASS detection (\citetalias{Xu2022}, \url{https://github.com/wwxu/rxgcc.github.io}).}
\label{table:RaDec}
\centering
{\def\arraystretch{1.2}\tabcolsep=3.5pt
\begin{tabular}{c c c c c c c}
\hline\hline
 RXGCC & Obs  & Obs ID & exp & RA & Dec & $z$ \\
& No. &  & [ks] & J2000 & J2000 & \\
\hline
  127 &   1  & 0863880101 & 15.7 & 46.011 & $-$12.073  &   0.012 \\
  841 &  2 & 0863880201 & 11.3 & 316.826 & $-$47.169 & 0.016 \\
  104 & 3 & 0863880401 & 18.5 & 36.488 & 36.996 & 0.036 \\
\hline
\end{tabular}
}
\end{table}
   
In Sect.~\ref{sect:2}, we discuss the XMM-Newton data reduction steps. We perform an imaging analysis (Sect.~\ref{sect:imaging}) followed by   surface brightness and spectral analysis in Sect.~\ref{sect:SB}. In Sect.~\ref{sect:results}, we present and discuss the results and in Sect.~\ref{sect:conclusion} we summarize the main conclusions of this work. 
Throughout this paper we assume a flat $\Lambda$CDM cosmology with $H_0 = 70$ km s Mpc$^{-1}$, $\Omega_{\rm M} = 0.3$ and $\Omega_\Lambda = 0.7$.

\section{Data reduction}\label{sect:2}
Data reduction and analysis is performed using HEASoft version 6.25 and SAS (Science Analysis Software) version 18.0.0 \texttt{(xmmsas\_20190531\_1155)}. The procedure is the same as described in \cite{Averonica} for the XMM-Newton data. Therefore, we provide only a short summary here.

As a first step we reduce and calibrate the event lists for the four targets observed with the XMM-Newton telescope in AO19 (P.I.: Thomas H. Reiprich). In particular, we calibrate photons in terms of energy and position in the sky and we perform a flare filtering operation which consists in identifying Good Time Intervals (GTIs) by discarding those that are contaminated by soft proton flares \citep[SPF,][]{DeLucaMolendi, KuntzSnowden}. A light curve and the corresponding count histogram were generated for each detector. A Poisson law was fitted to the histogram and time intervals falling outside $\pm3\sigma$ were rejected. In order to check for residual SPF contamination, we carried out the IN/OUT ratio test \citep{DeLucaMolendi, Leccardi} in the soft and hard X-ray band. A flare-free IN/OUT ratio should be close to unity. Unfortunately, Observation 4 failed the test with an IN/OUT ratio $>1.15$ and thus is not used for the analysis. In Observation 1, 2, and 3, contamination by SPF was well eliminated, leaving 10--20\,ks of good exposure time each.

Secondly, we checked the status of single chips in MOS1 and MOS2 since some of them might behave anomalously and show an unusually low hardness ratio and a particularly high background rate \citep{KuntzSnowden}. This resulted in MOS1-4 being removed in Observation 2. In addition, MOS1-3 and MOS1-6 were not in use during Observation 1, 2, and 3.
For the background treatment, we follow the same procedure as described in \cite{FWC} and \cite{Migkas2020}. We obtain background subtracted and exposure corrected images in the 0.4--1.25 keV energy band for Observation 1, 2, and 3, which enable a visual inspection of the targets.

The next step is the masking of point sources in the FoV, which can contaminate the morphological, photometric, and spectral analysis of the groups. This was done following the procedure described in \cite{Pacaud2006}: first a wavelet filtered image for the combined detectors in the 0.5--2 keV energy band was generated, then the Source Extractor software \citep[SExtractor,][]{SExtractor} was applied to detect point sources in the FoV and stored them in a catalog. After that, we checked the preliminarily cleaned images and manually added any missed point sources to the catalog. The manual inspection was done both in the hard (2--10 keV) and in the soft X-ray band (0.5--2 keV). This final catalog was then used in both imaging and spectral analyses to remove the unwanted sources.
Once the point sources were removed,
the area was refilled with random background photons for display purposes.

\section{Visual inspection and optical galaxy redshifts} \label{sect:imaging}
In this Section, we describe how we associate known galaxy clusters in our three XMM-Newton observations with detected extended X-ray sources.
For this, we make use of the RASS and XMM-Newton images shown in Fig.~\ref{fig:AllImages}.
Note that the extent (red circle in the RASS images), which represents the core radius from a maximum likelihood fitting of a beta model, is always smaller than the significance radius within which cluster emission was significantly detected, and also always smaller than $R_{500}$ within which the fluxes were determined in \citetalias{Xu2022}.
We also record the redshift, $z_{\rm assoc}$, of associated clusters. Furthermore, we determine optical redshifts for the extended X-ray sources, $z_{\rm opt}$, from known galaxy redshifts in the fields. 
We used the NED galaxy redshift information of the most likely members of the identified extended sources to calculate the median redshift. Additionally, we determined the median absolute deviation and scaled it to correspond to a $1\sigma$ Gaussian error. We take these values as their optical redshifts and uncertainties.
Later, in Section~\ref{sect:results}, we also provide X-ray redshifts, $z_\textrm{X-ray}$, when possible.
A summary of all redshifts for all groups and clusters, as well as the final adopted redshift, is provided in Table~\ref{table:results}.

\subsection{Observation 1 / RXGCC 127: Group 1 and Group 2}\label{subsec:Obs1}
In the XMM-Newton image of Observation 1 (Fig.~\ref{fig:XMM127}) one notices the emission from two bright extended X-ray sources close to the center of the pointing.
The northern object, which we call Group 1 from now on, could, in principle, be associated with a previously identified galaxy group: USGC S110 at $z \sim 0.012$ \citep{Mahdavi}; this redshift is also similar to the redshift associated with RXGCC 127 in \citetalias{Xu2022} ($z_{\rm assoc} = 0.012$). However, the central coordinates of USGC S110 (R.A., Dec.) = (46.0463, $-$12.0842) are far from the X-ray center of Group 1 ($6.9'$ offset) but instead are much closer to the X-ray center of the southern source, which we call Group 2 (angular separation of $1.9'$).

We found eight potential galaxy members of Group 1 in NED within the FoV (green dots in Fig.~\ref{fig:XMM127}), with redshifts ranging between 0.011 to 0.030. For Group 1, we obtained $z_\mathrm{opt}=0.013\pm0.002$. We note that NGC 1200 \citep[$z=0.0135$;][]{Wagner_2003} is located at the center of Group 1 and might be the BGG of the group.
Furthermore, we found 11 galaxies that potentially belong to Group 2 (cyan dots in Fig.~\ref{fig:XMM127}). The galaxies are located at $0.086\leq z \leq 0.136$. The median redshift is $z_\mathrm{opt}=0.102\pm0.014$. The WISEA J030405.18-120612.1 galaxy ($z=0.092$) coincides with the X-ray peak of Group 2.
Given that neither for Group 1 (no close coordinate match) nor for Group 2 (no close redshift match) a clear association with USGC S110 can be made, we keep their names.

\subsection{Observation 2 / RXGCC 841: Abell S0924, Group 3, Abell 3742, Abell 3745}
Observation 2 is shown in Fig.~\ref{fig:XMM841}. The presence of multiple extended sources is striking. On NED, three galaxy clusters are identified in the field, Abell S0924 at $z_{\rm assoc} \sim 0.016$ \citep{Coziol_2009}, Abell 3742 also at $z_{\rm assoc} \sim 0.016$ \citep{ACO_1989,Abell3742}, and Abell 3745 with unknown redshift.

For AS0924, we detect well-centered diffuse X-ray emission. For A3742 as well, albeit with an offset of about $2.5'$, (R.A., Dec.) = (316.674, $-$47.149); such an offset is not unusual given the large uncertainty of Abell cluster positions. Even modern optical cluster centers are often offset from X-ray centers; e.g., \citet{2023A&A...671A..57S} find an average offset of about 80 kpc, which corresponds to almost $4'$ at the redshift of A3742.
The third cluster, A3745, can also be associated with extended X-ray emission at the southern edge of the XMM-Newton FoV.

We found ten galaxies of known redshift in the range of $0.0162\leq z \leq 0.0476$, which could potentially belong to AS0924 or A3742. From this, we formally obtained $z_\mathrm{opt}=0.032\pm0.0183$. But note that the NGC 7014 galaxy ($z=0.0162$) sits at the center of AS0924, which is consistent with being the BCG of the cluster as reported in \cite{Coziol_2009}. \cite{Smith_2000} associated six galaxies with spec-\textit{z} in a large field to A3742, including ESO~286-G049 ($z=0.0177$) which coincides with our X-ray peak of A3742, and it seems reasonable to assume it to be the BCG.

Close to $0.5R_{500}$ of AS0924 (white dashed circle in Fig.~\ref{fig:XMM841}) and also to A3472 (magenta circle, centered on the X-ray peak), we identified another clump showing extended X-ray emission, and we refer to it as Group 3 (white circle). 
For Group 3, no galaxies with known redshifts are found in its vicinity.

For A3475, we find three galaxies of similar photometric redshift located within $\sim\!2.7'$ from the cluster center (yellow dots in Fig.~\ref{fig:XMM841}).
The median redshift of these galaxies is $z_{\rm opt}=0.130\pm0.003$.

\subsection{Observation 3 / RXGCC 104: Group 4, A0349, and [TKK2018] 859}\label{subsec:obs4}
In Observation 3, XMM-Newton data reveal the presence of three extended emission regions. One is close to the center of the pointing, with a slight offset towards the south-west. We associate this with the central part of RXGCC 104 (\citetalias{Xu2022}), and we call it Group~4 from now on. The reason for the center offset is likely due to the point source north-east of Group~4 that was not deblended in the RASS analysis (Fig.~\ref{fig:ROSAT104}).
We identified 16 galaxies in the field, possibly members of Group 4, for which a redshift estimate is available on NED spanning the range between $z \sim 0.028$ and $z \sim 0.042$, which if interpreted as comoving distance, corresponds to 119 Mpc and 178 Mpc. From this redshift information, we obtained $z_\mathrm{opt}=0.037\pm0.004$.
Situated at the center of Group 4 is the UGC 01877 galaxy (R.A., Dec =  36.409, 36.964). With $z=0.036$ \citep{Huchra_1999}, this galaxy is likely the BGG of the group.

Extended emission is also visible in the northern and southern outskirts of the FoV. The southern emission can be attributed to the galaxy cluster Abell 0349 \citep{ACO_1989}. The redshift of this cluster is unknown and it is quite challenging to conclude whether the identified galaxies in the field are members of the Group 4 or of Abell 0349. One galaxy of known redshift (WISEA J022620.10+364931.7; $z=0.128$) was found in NED, located near the center of A0349. Assuming that the cluster is at this redshift, the $\sim\!8'$ apparent extension translates to 1.1 Mpc. We also found three other galaxies of similar redshift to the redshift of WISEA J022620.10+364931.7 (yellow dots in Fig.~\ref{fig:XMM104}). Using these four galaxies, we obtained $z_\mathrm{opt}=0.136\pm0.017$.
One other galaxy of known redshift is also found in the perimeter of A0349, that is WISEA J022609.70+364704.2 located at (R.A., Dec.)\ = (36.540, 36.785) and $z=0.0322$ (green dot close to A0349 in Fig.~\ref{fig:XMM104}).

In addition, there is a low-significance detection of X-ray emission in the northern region, close to the edge of the FoV.
\citet{Tempel_2018} compiled a catalog of galaxy groups using a Bayesian group-finding technique based on marked point processes applied to the Two Micron All Sky Survey \citep[2MASS,][]{Skrutskie_2006} Redshift Survey \citep[2MRS,][]{Huchra_2012} data set. Using this catalog, we identified the [TKK2018] 859 galaxy group near the northern emission. This group is centered at (R.A., Dec.)\ = (36.387, 37.148), which corresponds to the center of the northern magenta circle in Fig.~\ref{fig:XMM104}. It has a spectroscopic redshift of 0.036, and it is associated with eight galaxies, four of which lie within the FoV. Since [TKK2018] 859 and Group 4 are at similar redshifts, it is challenging to distinguish, which galaxies within the FoV belong to which group. Therefore, we use the same $z_{\rm opt}=0.037$ estimate for both.
For completeness, \citet{USGCU118} list the optically-selected galaxy group USGC U118 at (R.A., Dec.)\ = (36.404, 37.123), just $1.7'$ away from [TKK2018] 859. The quoted redshift for this group, $z = 0.035$, is consistent with the one for [TKK2018] 859.
We also note that since the extended emission regions we associate with the A0349 cluster and the [TKK2018] 859 group are too close to the edge of the XMM-Newton FoV, we could not reliably determine their X-ray centers and, therefore, we reported their literature centers (Table~\ref{table:results}).

\section{X-ray analysis}
\label{sect:SB}
\label{sec:SpectralAnalysis}
The surface brightness analysis as well as the spectral analysis are performed on the masked images before the refilling operation.
For the surface brightness analysis, we have first performed a consistency check between the fluxes measured by the two instruments, ROSAT and XMM-Newton, by comparing the growth curve in \citetalias{Xu2018} with the one we determine from the XMM-Newton images. To compare them, we centered the profiles on the same point, i.e. the center of the pointing. The analysis is performed in the 0.5--2 keV energy range and the bin width used is $30''$. 

Then, we focus on individual groups observed in the XMM-Newton pointings. We build surface brightness profiles centering on the X-ray emission peak.
When needed, we mask the side of the source facing any nearby bright extended source (e.g., in the case of Group 1 and Group 2). Using \texttt{lmfit} \citep{lmfit}, we modelled the surface brightness profiles with a single $\beta$-model \citep{Cavaliere1976, Cavaliere1978}, as it generally provides a reasonable representation of the observed surface brightness for clusters and groups, and because we want to obtain a simple parameter describing the slope of the profiles. We also estimated the unabsorbed flux contribution. The latter step allows us to check whether the flux measured for each source is above or below the limit set for previous RASS-based cluster catalogs.

In this work, we performed spectral analysis using \texttt{XSPEC version 12.10.1} \citep{XSPEC}. 
The energy band that we have considered for the spectral analysis of all targets covers the range 0.5--7 keV. For each camera, we modeled and subtracted the instrumental background based on rescaled Filter-Wheel-Closed observations. In our initial modeling, we observed suspicious residual emission at high energy in most MOS1 spectra. Since, despite extensive checks, we could not identify any trace of residual particle flare contamination, nor any issue with the other cameras, we concluded that the residuals resulted from inconsistencies between our templates for the MOS1 instrumental background and the real data. We, therefore, excluded MOS1 from subsequent spectral analyses.
This reduction of $\sim$15\% of effective exposure time does not affect the precision of our results significantly.

The spectral model that is fit to the data has two terms; i.e.,
\bigskip
\noindent
$\texttt{Model} = \texttt{constant} (\texttt{apec}_1 + \texttt{phabs} (\texttt{apec}_2 + \texttt{powerlaw}) + \texttt{gaussian}_1 + \texttt{gaussian}_2 ) + \texttt{phabs} \times \texttt{apec}_3$.
\bigskip
\noindent
The first term is used to fit the cosmic X-ray background (CXB) and the residuals of the instrumental lines, rescaled to the area of the regions employed for spectral extraction in arcmin$^2$ (\texttt{constant}). The first apec model, $\texttt{apec}_1$, accounts for the emission coming from the Local Hot Bubble (LHB), while $\texttt{apec}_2$ is used to model the Milky Way halo (MWH). The latter, together with the \texttt{powerlaw} model which describes the unresolved AGN, are multiplied by $\texttt{phabs}$ to take into account the local absorption. The total hydrogen column density, which includes neutral and molecular hydrogen, is given by \cite{nH}.

The Gaussian components that were included in the model account for emission from the main fluorescence lines present in the instrumental background. Indeed, these are time variable, and possible residuals can remain in the background subtracted spectra. For the MOS2 detector, these include Al K$_\alpha$ at 1.4865 keV and Si K$_\alpha$ at 1.74 keV. For PN, only Al K$_\alpha$ is modeled (while the normalization of the other line is fixed to zero). During spectral fitting, the energies and (zero) widths of the emission lines were fixed, while their normalization was free to vary.

Finally, the second term, $\texttt{phabs} \times \texttt{apec}_3$, is used to model the absorbed cluster emission. Unless specified otherwise, temperature, metallicity and normalization of $\texttt{apec}_3$ are linked between MOS2 and pn and are free to vary during the spectral fitting in order to estimate these quantities for each cluster or group in our sample.

\begin{table}
\centering
\caption{Starting values for the parameters describing the CXB in our spectral model.}
\resizebox{\columnwidth}{!}
{\begin{tabular}{c|c c c }
\hline\hline
Component& Parameter & Value & Fixed/Variable \\
\hline
$\texttt{apec}_1$ &  $k_\mathrm{B}T$ [keV] &  $0.099^{a}$  & fixed \\
    & $Z \, [Z_\odot]$ & $1$  & fixed \\
    & $z$   & $0$  & fixed \\
    & $norm \, [\mathrm{cm}^{-5}/\mathrm{arcmin}^2]$ & $1.7 \times 10^{-6}$ & variable\\
\hline
    $\texttt{phabs}$ & $n_\mathrm{H}$ [atoms cm$^{-2}$] & $7.49 \times 10^{20}$ & fixed \\
\hline
$\texttt{apec}_2$ & $k_\mathrm{B}T$ [keV] & $0.225^{a}$ & fixed \\
    & $Z \, [Z_\odot]$ & $1$    & fixed \\
    & $z$ & $0$    & fixed \\
    & $norm \, [\mathrm{cm}^{-5}/\mathrm{arcmin}^2]$ & $7.3\times 10^{-7}$    & variable \\
\hline
$\texttt{powerlaw}$ & $\Gamma$ & $1.45^{b}$   & fixed \\
    & $norm^\ast$ & $5 \times 10^{-7}$ & variable \\
\hline
\multicolumn{4}{l}{\footnotesize $^\star$ in $[\mathrm{photons/keV/cm^2/s/arcmin^2}$] at 1 keV, {$^a$\cite{McCammon_2002}}, }\\
\multicolumn{4}{l}{\footnotesize {$^b$\cite{DeLucaMolendi}}}\\
\end{tabular}}
\label{table:model}
\end{table}

For each observation, we need to model the sky background. To do so, the spectrum from a region where cluster emission does not dominate is required. Since we expect the studied galaxy groups to be rather extended, sky background spectra were extracted from ROSAT data in the surroundings of the XMM-Newton pointings. In particular, the same annular regions used in \citetalias{Xu2018} for the sky background estimate were considered.

The number of X-ray clumps revealed by our XMM-Newton observations coinciding with the RASS detections in \citetalias{Xu2022} adds some ambiguity in the X-ray to optical association process and redshift determination. In addition, Groups 1--4 could not be clearly associated with a previously known optical group. Therefore, whenever possible, we also derived an X-ray redshift for each source from our spectral fits. Usually, this required to extract a spectrum for the higher signal-to-noise core emission (the exact aperture was estimated on a cluster-by-cluster basis based on visual inspection).
For Group 1 and 2 we used circular regions with $1'$ radius centered on their X-ray peak. In each case, half of the aperture was masked to exclude contamination from the other group.
For Group 3 the circular region from which we extracted the spectrum needed to be larger in order to have enough statistics to fit a redshift value from the source spectrum. With a radius of $1.6'$ centered on R.A. $=316.761$, Dec.\ $=-47.140$, we encompass most of the group emission and avoid contamination from the surrounding clusters.

Since the redshift of A0349 in Observation 3 was not available in the literature, we obtain an X-ray redshift estimate for this cluster as well, by extracting and fitting a spectrum from a circular region of radius $2'$ centered at (R.A., Dec.)\ = (36.500, 37.005). The spectrum extraction region was selected in such a way that we maximize the signal-to-noise ratio and avoid contamination from Group 4.
Since the cluster is only partially covered and may not be properly centered, we do not report the measured temperature.

We estimate the size of each cluster by deriving $R_{500}$ iteratively using core-excised temperatures within 0.2--0.5\,$R_{500}$ and applying the \citet{Lovisari} $M_{500}$--$T$ relation.
To estimate errors of the resulting $R_{500}$, we take into account the statistical uncertainty of the temperature but neglect the intrinsic scatter of the $M_{500}$--$T$ relation as well as uncertainties of the average relation.
Emission from nearby groups was masked in the process. For AS0924, we excluded the entire western half for this reason. Furthermore, 0.5 $R_{500}$ partially exceeds the usable XMM-Newton FoV for AS0924, resulting in the estimated temperature being weighted more towards the inner regions.

Group 3 represents a special case due to low counts and vicinity to other sources. Therefore, we iteratively used the surface brightness profile to determine $L_{X}$ and employed the $L_{X}$--$M_{500}$ relation from \citet{Lovisari} to derive $R_{500}$. As a consistency check, this alternative procedure was also applied to the other groups and yielded similar results within a large scatter. Due to the low number of counts, for Group~3, the reported temperature was determined within a region of $1.6'$ radius.

\section{Results and discussion}
\label{sect:results}

We summarize the results of the analysis carried out in this work in Table~\ref{table:results}.
The table is divided into three parts, corresponding to the three XMM-Newton observations of RXGCC 127, 841, and 104. The redshifts determined by \citetalias{Xu2022} are also provided in the header rows.
Column~1 contains the names of all nine galaxy groups. The coordinates (Columns~2 and 3) for Groups 1--4 are the X-ray centers as determined with the XMM-Newton observations here. This is also true for A3742 and AS0924. For the other groups, the literature coordinates are given because they are too close to the edge of the XMM-Newton FoV.
The next two columns contain the redshifts determined in Sect.~\ref{sect:imaging}, $z_{\rm opt}$ and $z_{\rm assoc}$. In Column~6, we list the redshift obtained from X-ray spectral fitting (Sect.~\ref{sec:SpectralAnalysis}); further details on $z_\textrm{X-ray}$, as well as on the finally adopted redshift (indicated in Column~7) are provided below.
Columns~8--11 show the determined X-ray temperatures, $M_{500}$, $R_{500}$ (both in arcmin and kpc; see Sect.~\ref{sec:SpectralAnalysis} for how they were determined).
The best-fit $\beta$ values from the surface brightness profile analysis (Sect.~\ref{sect:SB}) are given in Column~12.
Finally, background-subtracted, unabsorbed fluxes measured in two energy bands are provided in Columns~13 and 14.

In the observation of RXGCC 127 ($z_{\textrm{X22}} = 0.012$), we clearly detect two very distinct galaxy groups, called Group 1 and Group 2. The redshift analyses in Sects.~\ref{sect:imaging} and \ref{sec:SpectralAnalysis} reveal that these two groups reside at two very different redshifts and are, therefore, not physically connected.
For Group~1, we determined $z_{\textrm{opt}}=0.013\pm0.002$, which coincides almost exactly with the redshift determined by \citetalias{Xu2022}, $z_\textrm{assoc}$. From the X-ray spectral analysis, we found $z_\textrm{X-ray}=0.020_{-0.008}^{+0.014}$, which is also consistent. We, therefore, adopt $z_\textrm{opt}=0.013$ for Group 1.
We found that Group 2 is at higher redshift, with consistent optical and X-ray redshifts, $z_\textrm{opt}=0.102\pm0.014$ and $z_\textrm{X-ray}=0.107^{+0.014}_{-0.008}$, and we adopt the latter.

From its temperature, $k_\textrm{B}T=0.84\pm0.06$\,keV, we derived that Group 1 is a low mass group with $M_{500}$ just above $10^{13}\,M_\odot$ (Table~\ref{table:results}). Its surface brightness profile is flatter than the canonical $\beta=2/3$ but not by much. On the other hand, Group 2 is hotter, $k_\textrm{B}T=3.03^{+0.40}_{-0.36}$\,keV, and, therefore, more massive, just above $10^{14}\,M_\odot$, as expected given the higher redshift. Its surface brightness profile, $\beta=0.53\pm0.03$, is significantly flatter than 2/3 but still above 1/2.

The XMM-Newton observation of RXGCC 841 ($z_{\textrm{X22}} = 0.016$) reveals that it is composed of four galaxy groups exhibiting extended X-ray emission, AS0924, Group 3, A3742, and A3745.
While the X-ray and optical redshifts of AS0924 are rather close to each other ($z_\textrm{X-ray}=0.027_{-0.014}^{+0.022}$; $z_{\rm opt}=0.032\pm0.018$), we decided to adopt $z_{\rm assoc}=0.016$ as the final redshift for this system because of the redshift of the central galaxy (NGC 7014, $z = 0.0162$), which is anyway within $1\sigma$ of both $z_\textrm{X-ray}$ and $z_{\rm opt}$. Since for Group 3 we only have the X-ray redshift, $z_\textrm{X-ray}=0.235^{+0.048}_{-0.054}$, we adopt it. The X-ray signals of A3742 and A3745 are too faint to estimate X-ray parameters.
We adopt $z_{\rm assoc}=0.018$ for A3742 because of the co-incidence of galaxy and X-ray peak. This redshift is also consistent within $1\sigma$ with $z_{\rm opt}=0.032\pm0.018$. For A3745, we adopt $z_{\rm opt}=0.130\pm0.003$.

We find low temperatures for both AS0924 and Group 3, $1.04^{+0.30}_{-0.25}$\,keV and $1.62^{+0.35}_{-0.15}$\,keV, respectively. This results in group-like masses of a few times $10^{13}\,M_\odot$ (Table~\ref{table:results}). With $\beta=0.53\pm0.05$ the surface brightness slope for Group 3 is similarly on the flat side as the ones of Group 1 and 2. Even more striking, however, is the $\beta=0.36\pm0.02$ for AS0924, which is quite unusual. For example, in the $\sim$100 bright nearby groups and clusters in HIFLUGCS \citep{Reiprich2002}, not a single system has such a small value of $\beta$, although low $\beta$ values are not unheard of for some of the lowest mass groups and elliptical galaxies \citep[e.g.,][]{2003MNRAS.340..989S}.

In the observation of RXGCC 104 ($z_{\textrm{X22}} = 0.036$) , we detect three extended X-ray emission regions, Group~4, A0349, and [TKK2018] 859. For the centrally located Group~4, all three redshift estimates, $z_\textrm{opt}$, $z_\textrm{X-ray}$, $z_\textrm{assoc}$, are consistent with each other (Table~\ref{table:results}), and we adopt $z_\textrm{assoc}=0.036$ from \citetalias{Xu2022}. For A0349, no associated redshift is available but $z_\textrm{opt}$ and $z_\textrm{X-ray}$ are consistent and we adopt $z_\textrm{X-ray}=0.139^{+0.017}_{-0.009}$.
As described in Sect.~\ref{sec:SpectralAnalysis}, since A0349 is only partially covered and may not be properly centered, we do not report a measured temperature.
No X-ray parameters could be determined for [TKK2018] 859 but our optical redshift estimate, $z_{\rm opt}=0.037\pm0.004$, is consistent with $z_{\rm assoc}=0.036$ from \citet{Tempel_2018} and we adopt the latter.

Our temperature estimate of $\sim$$1.5$\, keV for Group~4 results in $M_{500}\approx 5\times10^{13}\,M_\odot$. Also Group~4 exhibits a strikingly flat surface brightness profile, $\beta=0.39\pm0.01$.

\begin{table*}
\begin{center}
\caption{Summary of the galaxy group and cluster properties. See Sect.~\ref{sect:results} for details.}
\label{table:results}
\begin{adjustbox}{max width=\textwidth}

\begin{tabular}{ m{2.5cm} m{2cm}<{\centering} m{2cm}<{\centering} m{2.0cm}<{\centering} m{1.7cm}<{\centering} m{1.7cm}<{\centering} m{1.6cm}<{\centering} m{1.6cm}<{\centering} m{1.6cm}<{\centering} m{1.6cm}<{\centering} m{1.8cm}<{\centering} m{1.8cm}<{\centering} m{1.8cm}<{\centering} m{1.8cm}<{\centering} }
\hline\hline\\[-1ex]
\multirow{2}{*}{Galaxy group} & R.A.\ (J2000) & Dec.\ (J2000) & \multirow{2}{*}{$z_\mathrm{opt}$} & \multirow{2}{*}{$z_\mathrm{assoc}$} &\multirow{2}{*}{$z_\textrm{X-ray}$} & Redshift & $k_\mathrm{B}T$ & $M_{500}$ & $R_{500}$ & $R_{500}$  & \multirow{2}{*}{$\beta$} & $F_{0.5-2.0~\mathrm{keV}}^\ddagger$ & $F_{0.1-2.4~\mathrm{keV}}^\ddagger$\\ 
  &  [deg] & [deg] &  & &  & adopted &[keV] & [$10^{13}M_{\odot}]$ & [arcmin] & [kpc] & & \rule{0pt}{2.5ex}[$10^{-13}~\mathrm{\frac{erg}{s\,cm}}$] & \rule{0pt}{2.5ex}[$10^{-13}~\mathrm{\frac{erg}{s\,cm}}$]\\[1ex] 
\hline
\hline\\[-1ex]
\multicolumn{14}{c}{
RXGCC~127\ \
 ---\ \ $z_\mathrm{X22}=0.012$} \\[5pt]
\hline\\[-1.7ex]
Group 1 & $45.977$ & $-11.991$ & $0.013\pm0.002$ &
$0.012^\textrm{a}$ & $0.020_{-0.008}^{+0.014}$ & opt & $0.84^{+0.06}_{-0.06}$ & $1.86^{+0.25}_{-0.25}$ & $25.15^{+1.15}_{-1.13}$ & $400.85^{+18.32}_{-18.08}$ & $0.58\pm0.03$ & $4.37\pm0.36$ & $7.74\pm0.63$ \\[5pt]
Group 2 & $46.021$ & $-12.103$ & $0.102\pm0.014$ &
-- & $0.111_{-0.008}^{+0.013}$ & X-ray& $2.80_{-0.28}^{+0.30}$ & $13.48_{-2.31}^{+2.47}$ & $6.20_{-0.35}^{+0.38}$ & $752.33_{-42.95}^{+45.92}$ & $0.53\pm0.03$ & $3.81\pm0.12$ & $7.25\pm0.23$ \\[5pt]
\hline\\[-1ex]
\multicolumn{14}{c}{
RXGCC~841\ \
 ---\ \  $z_\mathrm{X22}=0.016$} \\[5pt]
\hline\\[-1.7ex]
AS0924 & $316.968$ & $-47.178$ & $0.032\pm0.018$ &
$0.016^\textrm{b}$ & $0.027_{-0.014}^{+0.022}$ & assoc & $1.04^{+0.30}_{-0.25}$ & $2.63^{+1.28}_{-1.07}$ & $23.01_{-3.12}^{+3.73}$ & $449.88_{-61.00}^{+72.85}$ & $0.36\pm0.02$ & $1.02\pm0.13$ & $1.68\pm0.22$ \\[5pt]
A3742 & $316.699$ & $-47.188$ & $0.032\pm0.018$ & $0.018^\textrm{c}$ & Unconstr. & assoc & -- & -- & -- & --  & --  & -- & -- \\[5pt]
A3745 & $316.925$ & $-47.415$ & $0.130\pm0.003$ & -- & Unconstr. & opt & -- & -- & -- & -- & -- & -- & -- \\[5pt]
Group 3 & $316.768$ & $-47.137$ & -- & -- & $0.235_{-0.054}^{+0.048}$ & X-ray & $1.62_{-0.15}^{+0.35}$ & $4.04^{+0.65}_{-0.74}$ & $2.15_{-0.12}^{+0.12}$ & $482.09^{+27.91}_{-27.50}$ & $0.53\pm0.05$ & $0.38\pm0.05$ & $0.67\pm0.09$ \\[5pt]
\hline\\[-1ex]
\multicolumn{14}{c}{
RXGCC~104\ \
 ---\ \ $z_\mathrm{X22}=0.036$} \\[5pt]
\hline\\[-1.7ex]
Group 4 & $36.409$ & $36.965$ & $0.037\pm0.004$ & $0.036^\textrm{a}$ & $0.031_{-0.003}^{+0.003}$ & assoc & $1.49^{+0.05}_{-0.05}$ & $4.77^{+0.37}_{-0.37}$ & $12.70^{+0.33}_{-0.33}$ & $545.32^{+14.00}_{-14.13}$ & $0.39\pm0.01$ & $9.66\pm0.47$ & $17.43\pm0.85$ \\[5pt]
A0349 &  $36.600$ & $36.817$ & $0.136\pm0.017$ & -- & $0.139_{-0.009}^{+0.017}$ & X-ray & -- & -- & -- & -- & -- & -- & -- \\[5pt]
[TKK2018] 859 & $36.387$ & $37.148$ & $0.037\pm0.004$ & $0.036^\textrm{d}$ & Unconstr. & assoc & -- & -- & -- & -- & -- & -- & -- \\[5pt]
\hline\hline\\[-1.5ex]
\multicolumn{14}{l}{\footnotesize $^\textrm{a}$\citetalias{Xu2022}, $^\textrm{b}$\citet{Coziol_2009}, $^\textrm{c}$ESO 286-G049 in \citet{Smith_2000}, $^\textrm{d}$\citet{Tempel_2018}}\\
\hline\hline
\end{tabular}
\end{adjustbox}
\end{center}
\end{table*}

\begin{figure}[ht!]
    \centering
    \includegraphics[scale=0.4]{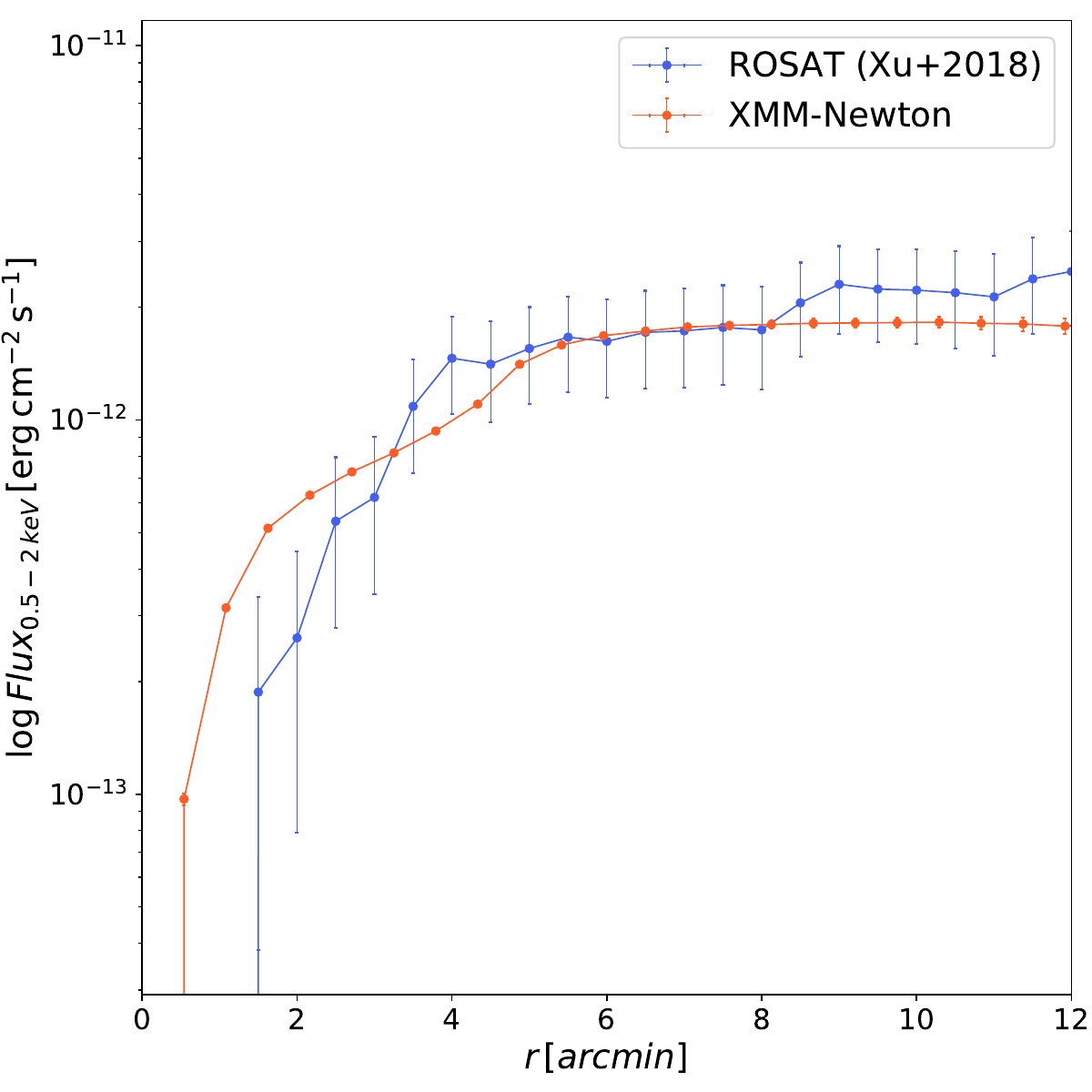} 
    \caption{Flux comparison performed between ROSAT and XMM-Newton for the observation of RXGCC 127 in the 0.5--2 keV energy range. The integrated-flux profiles are background subtracted.}
    \label{fig:Obs1_fluxcomparison}
\end{figure}

We now turn to the flux measurements of the galaxy groups.
First, Fig.~\ref{fig:Obs1_fluxcomparison} shows the result of the flux comparison between ROSAT and XMM-Newton in the case of the observation of RXGCC 127. One can see that the flux measurements are consistent at all radii, except possible within the inner $\sim$$2'$, where the broader ROSAT PSPC PSF distributes some flux from inner to outer regions. From this test we conclude that the two instruments measure the same flux within the usable XMM-Newton FoV; i.e., within $12'$.

Based on the RASS observation, \citetalias{Xu2022} had associated the measured flux of RXGCC 127 to a single galaxy group at $z=0.012$. The much higher quality XMM-Newton image shows that the flux contribution instead comes from two distinct extended sources at different redshifts, namely Group 1 and Group 2. Similar cases have been presented in \citet{FWC} and \citet{2004ApJ...608..179R}, where follow-up XMM-Newton observations had revealed that the limited RASS spatial resolution ($\sim$ 1 arcmin) and photon statistics had caused some double or triple clusters, or unrelated clusters with small projected separation, to be identified as a single source.

\begin{figure*}
\centering
\subfloat{\includegraphics[width=0.45\textwidth]{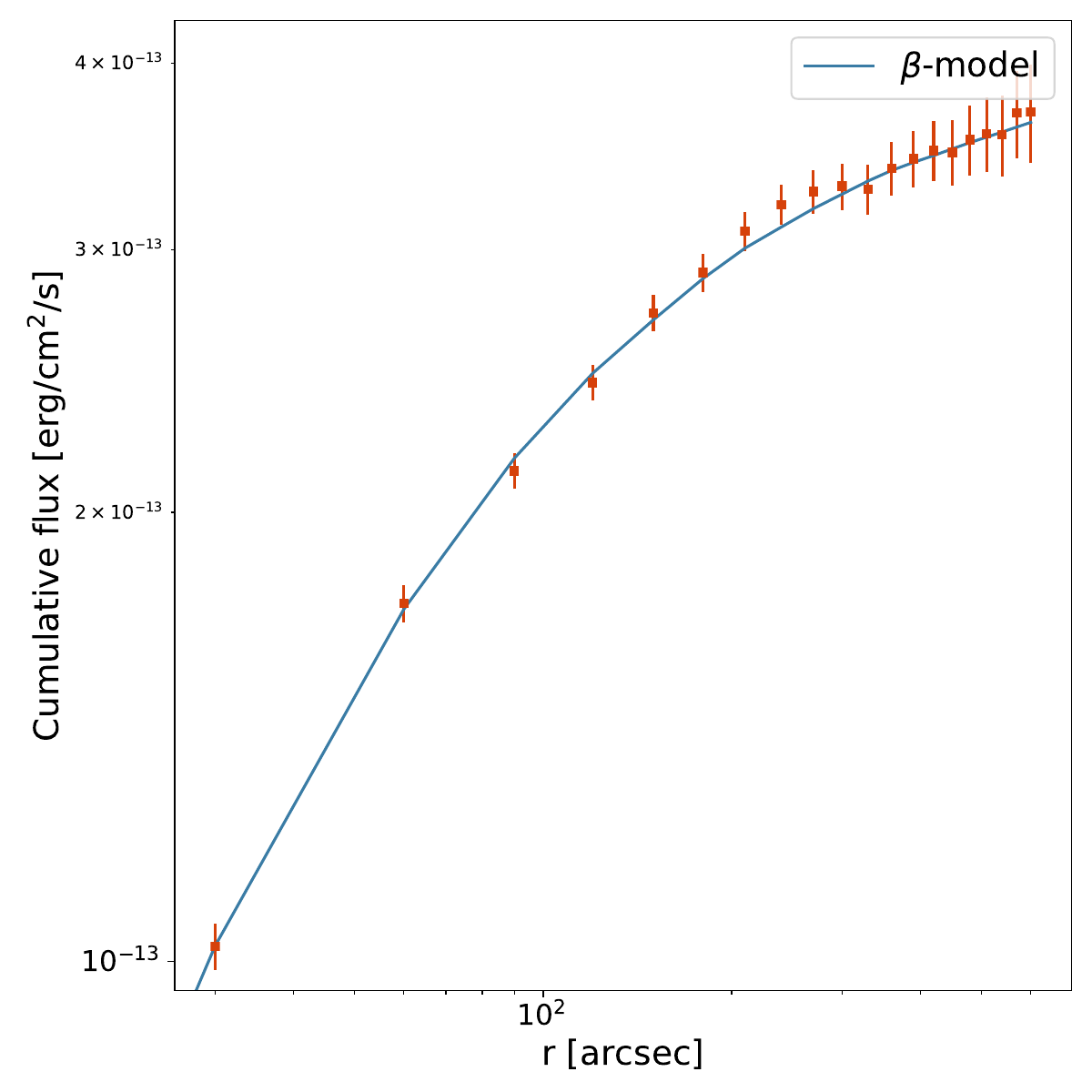}\label{fig:grp1_north_cumflux}}\hskip1ex
\subfloat{\includegraphics[width=0.45\textwidth]{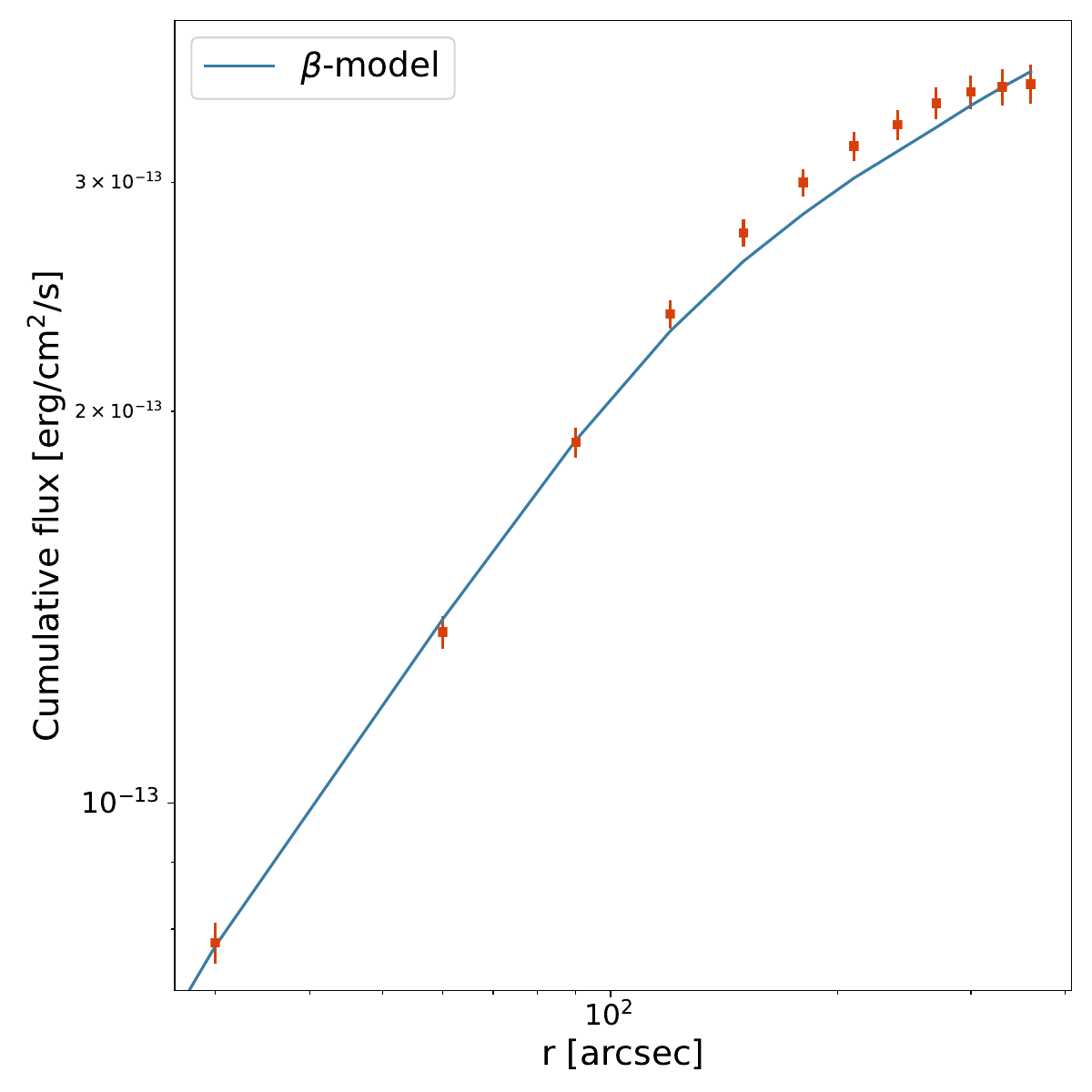}\label{fig:grp1_south_cumflux}}
\caption{Cumulative sky background subtracted flux in the 0.5-2 keV energy band and best-fit $\beta$-model for Group 1 (left) and Group 2 (right).}
\label{fig:grp1_individualfluxes}
\end{figure*}

Since XMM-Newton allowed us to distinguish the two sources from which the ROSAT flux is measured, we are interested in estimating what is the contribution of each of them to the total. The surface brightness profiles were built as described in Sect.~\ref{sect:SB}. Then we integrated the profiles, which are shown in Fig.~\ref{fig:grp1_individualfluxes} for Group 1 and Group 2. We integrated up to the radius at which the sky background emission starts to become dominant. Integrated flux profiles, and their best-fit $\beta$-models, for AS0924, Group 3, and Group 4 can be found in Appendix \ref{appendixB}.

In Table~\ref{table:results} we list all our flux measurements along with the best-fit $\beta$ values. The estimated fluxes all lie below the typical flux limit of previous RASS based galaxy cluster surveys (i.e., $\approx 3 \times$ 10$^{12}$ erg s$^{-1}$ cm$^{-2}$ in the $0.1-2.4$ keV energy band). Given our finding that the RASS-measured fluxes need to be split up into two (RXGCC 127), three (RXGCC 104), or even four (RXGCC 841) individual galaxy groups, this is not surprising. In some sense, this is good news for the previous RASS-based cluster surveys, as the systems characterized here do not make it above their flux limits. On the other hand, our results show again that projection effects can also affect X-ray selected cluster samples. Moreover, the groups identified here exhibit rather flat X-ray surface brightness profiles. In particular, the two main central groups, AS0924 and Group~4, have $\beta<0.4$. This demonstrates the need to account for such systems, as well as for projection effects, when studying the selection effects in X-ray cluster surveys. 

\section{Summary of conclusions}
\label{sect:conclusion}

We performed X-ray imaging,  surface brightness, and spectral analysis of an XMM-Newton subsample of galaxy groups found during a reanalysis of the RASS (\citetalias{Xu2018}; \citetalias{Xu2022}). This reanalysis had revealed the presence of several extended X-ray sources missed in the past, despite their measured fluxes being similar to the flux limits applied in previous RASS cluster catalogs. Some of these sources did not even have any X-ray detection before in any X-ray catalog. With our XMM-Newton analysis of a subsample, we wanted to test whether these sources might have been overlooked in the past because they exhibit very extended, flat surface brightness profiles for which some of the major source detection algorithms are not optimized.

The answer, based on this limited subsample, is yes and no. Our main results and conclusions are the following.
\begin{itemize}
    \item The XMM-Newton observations indeed reveal that the central groups (AS0924 and Group~4) of two out of the three analyzed observations (RXGCC 104 and RXGCC 841) have a very flat surface brightness profile with a slope parameter $\beta<0.4$.
    \item The RASS flux of these two observations measured by \citetalias{Xu2022} is decomposed into flux contributions from altogether seven individual groups.
    Out of these additional five groups, two (A3742 and [TKK2018] 859) appear to lie at the same distance as the respective main central group, while three (A3745, Group~3, and A0349) seem to just happen to lie projected along the line of sight.
    \item The third observation, pointed at RXGCC 127, reveals two groups with similar flux each (Group~1 and Group~2). These two groups are shown to lie at different distances, so are not physically connected. With $\beta<0.6$ they both also show a flat surface brightness profile but not extremely flat.
    \item The individual fluxes of these nine groups with extended X-ray emission all fall below the typical RASS cluster survey flux limits. So, strictly speaking, they are not missing from previous RASS-based cluster catalogs.
    \item Nonetheless, the main central groups do feature unusually flat surface brightness profiles, so the existence of such groups needs to be taken into account when determining selection functions; in particular for surveys with a source detection process optimized for point sources.
    \item Also, our analysis demonstrates that the vicinity of other groups at the same or a different redshift can affect a resulting cluster catalog, since several sources may be blended and detected as a single source in a cluster detection process with few photons and moderate spatial resolution. For example, if the goal was to constrain the redshift-dependent cluster mass function, not accounting for such a source type would result in an underestimate of the number density and, probably, an overestimate of the mass, and, possibly, an incorrect redshift.
    \item Our conclusions are based on a small set of follow-up observations of systems discovered by \citetalias{Xu2018} and \citetalias{Xu2022} and may, therefore, not be fully representative. Therefore, obtaining further XMM-Newton follow-up observations to corroborate the findings here would be useful. A representative subsample of about 25 systems in total from \citeauthor{Xu2022} should suffice. Extrapolating from the current study (four observations performed with average total observing time of about 17 ks each and three observations out of them fulfilling the quality criteria) a reasonable XMM-Newton investment of about 500 ks would be required, which appears feasible.
    Moreover, for groups at high redshifts, the future NewAthena X-ray observatory will be able to detect and characterize such systems \citep{2013arXiv1306.2319P,2025NatAs...9...36C}.
\end{itemize}

\begin{acknowledgements}
CS, AV, and TR acknowledge support from the German Federal Ministry of Economics and Technology (BMWi) provided through the German Space Agency (DLR) under project 50 OR 2112. AV acknowledges funding by the Deutsche Forschungsgemeinschaft (DFG, German Research Foundation) -- 450861021.
WX thanks the support of National Nature Science Foundation of China (Nos.\ 11988101, 12022306, 12203063), the support by National Key R$\&$D Program of China No.\ 2022YFF0503403, the support from the Ministry of Science and Technology of China (No.\ 2020SKA0110100), the science research grants from the China Manned Space Project (Nos.\ CMS-CSST-2021-B01, CMS-CSST-2021-A01), CAS Project for Young Scientists in Basic Research (No.\ YSBR-062), and the support from K.C.Wong Education Foundation.
KM acknowledges support in the form of the X-ray Oort Fellowship at Leiden Observatory.
Based on observations obtained with XMM-Newton, an ESA science mission with instruments and contributions directly funded by ESA Member States and NASA.
\end{acknowledgements}

\bibliographystyle{aa}
\bibliography{bibliography}

\appendix 

\onecolumn
\newpage
\section{Surface brightness profiles} \label{appendixB}

\begin{figure*}[h]
\centering
\vspace{1cm}
\subfloat{\includegraphics[width=0.45\textwidth]{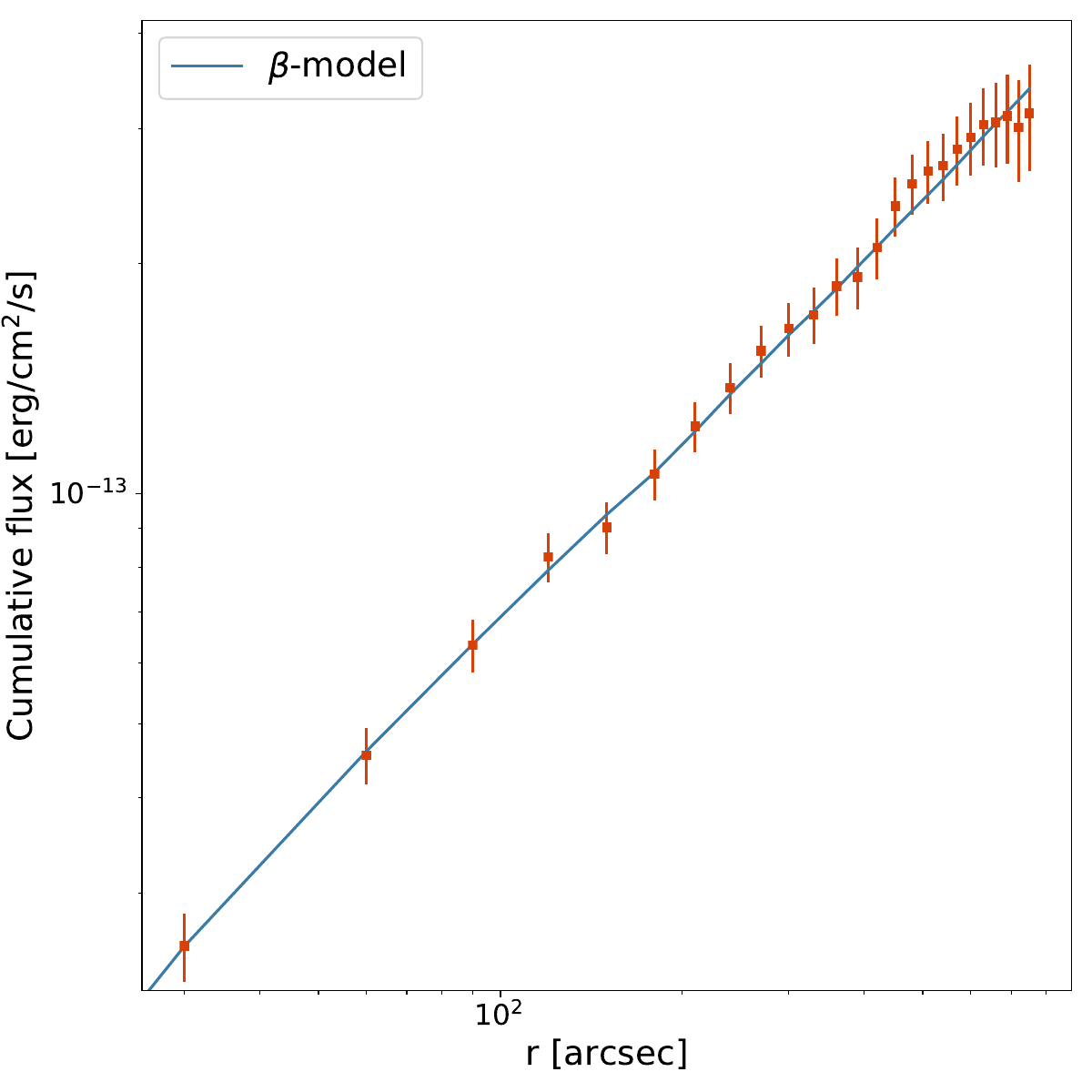}\label{fig:AS0924_FluxCUM}}\hskip1ex
\subfloat{\includegraphics[width=0.45\textwidth]{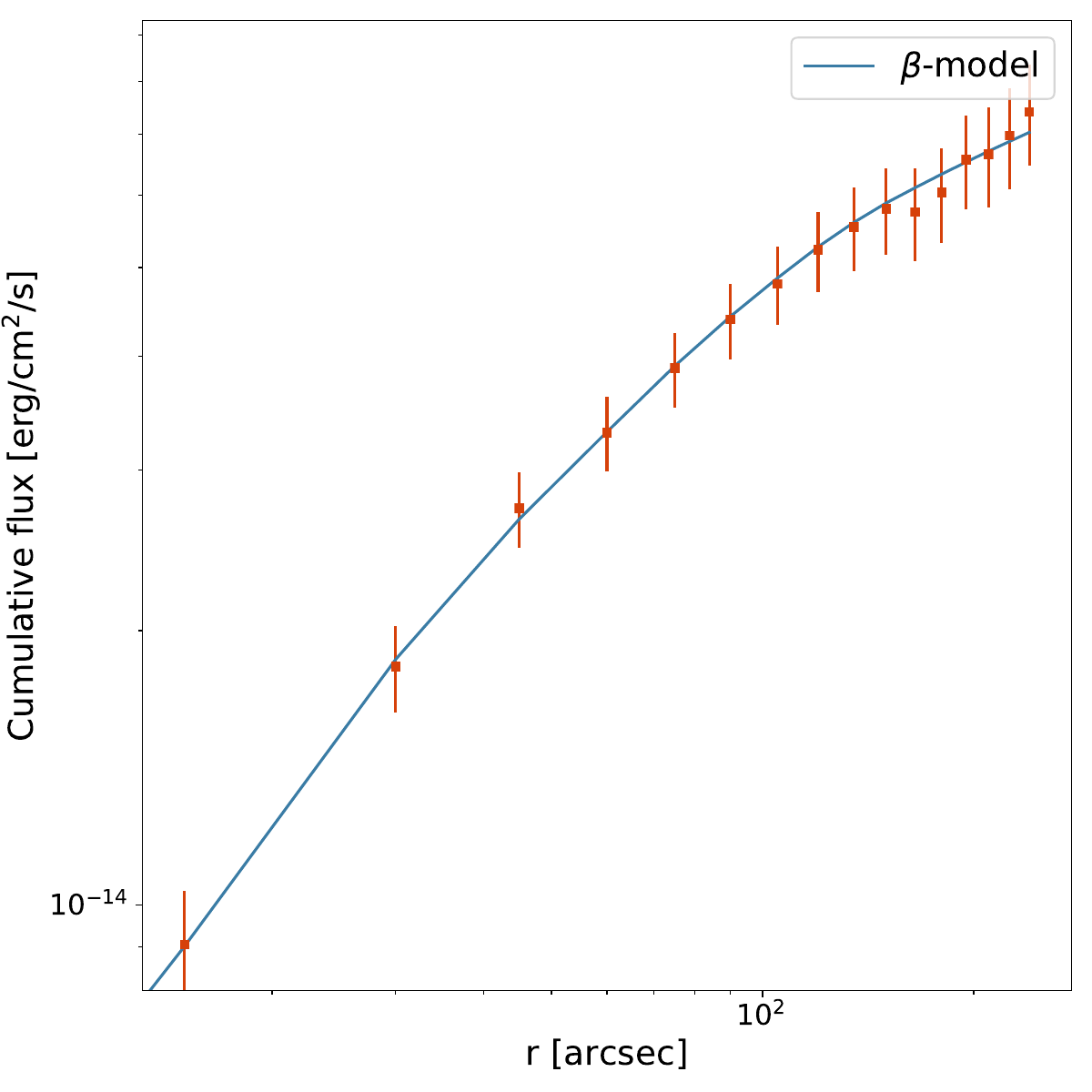}\label{fig:Group3_cumflux}}\hskip1ex
\subfloat{\includegraphics[width=0.45\textwidth]{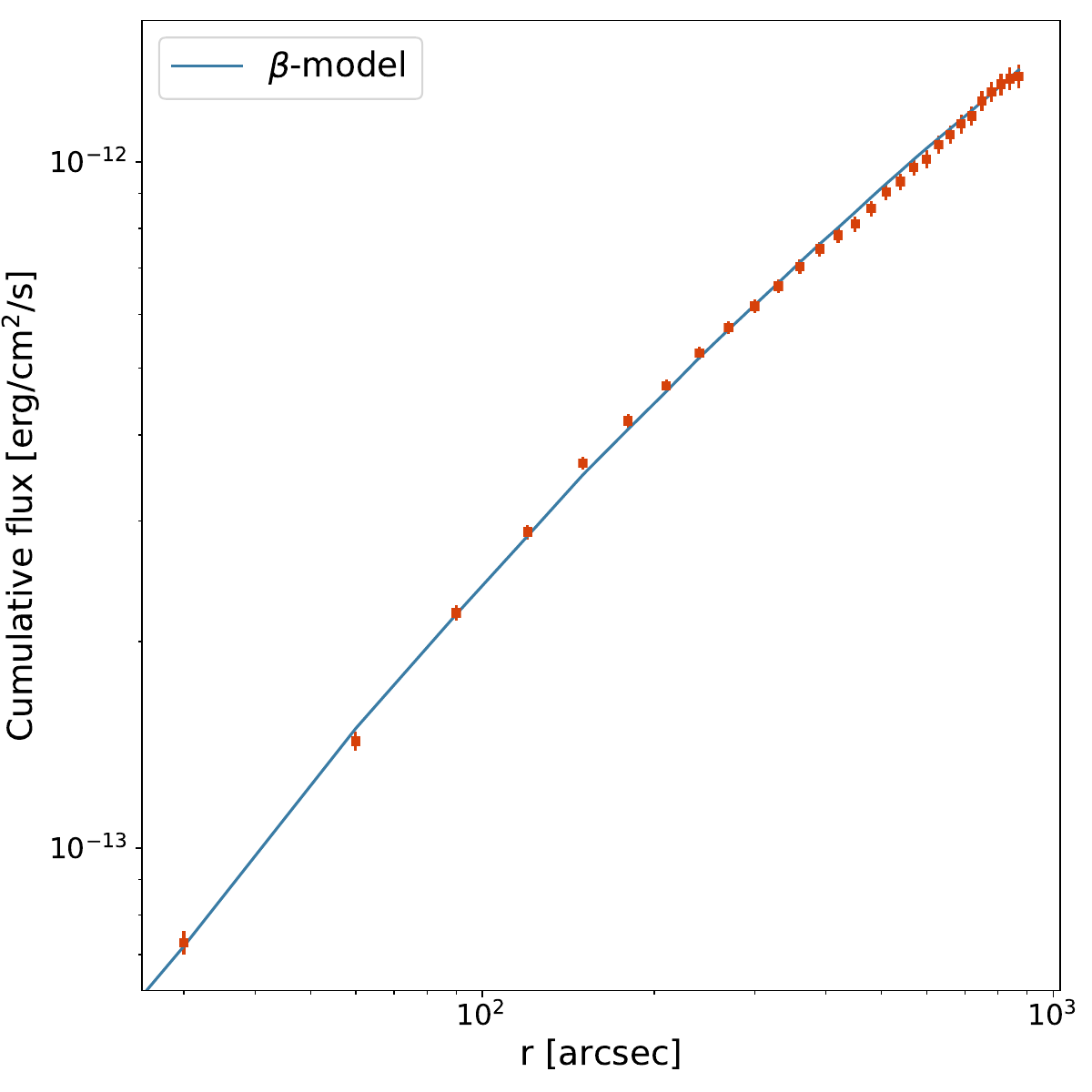}\label{fig:Group4_cumflux}}
\caption{Cumulative sky background subtracted flux in the 0.5-2 keV energy band and best-fit $\beta$-model for AS0924 (top left panel), Group 3 (top right panel) and Group 4 (bottom panel).}
\label{fig:grp4_individualfluxes}
\end{figure*} 

\newpage
\section{Extracted spectra} \label{app:Fx4.37Spectra}

\begin{figure*}[h]
\centering
\vspace{1cm}
\subfloat[Group~1 within $0.2$--$0.5$ $R_{500}$]{\includegraphics[width=9cm]{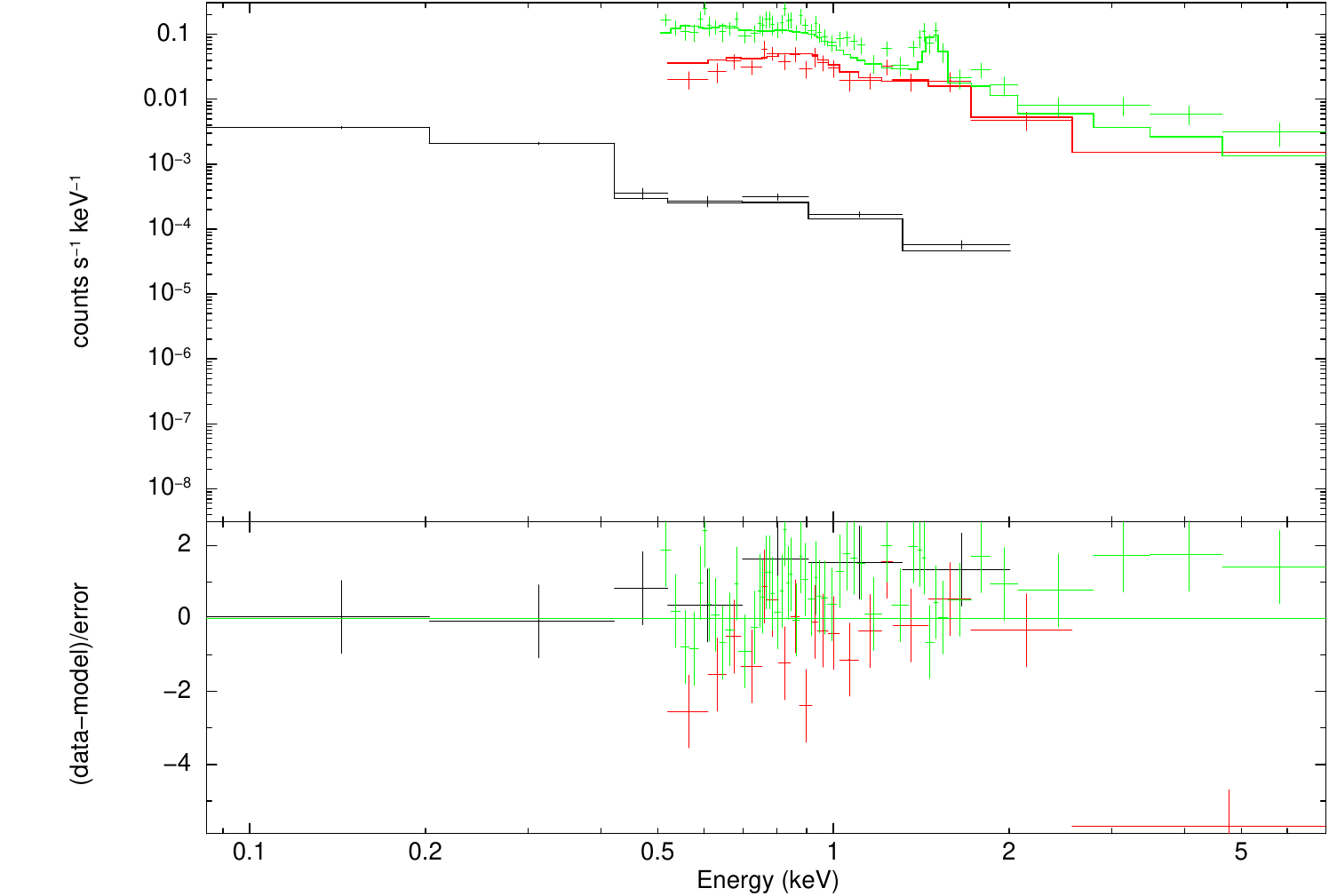}\label{fig:Group1Spec}}
\subfloat[Group~1 for $z$ estimation]{\includegraphics[width=9cm]{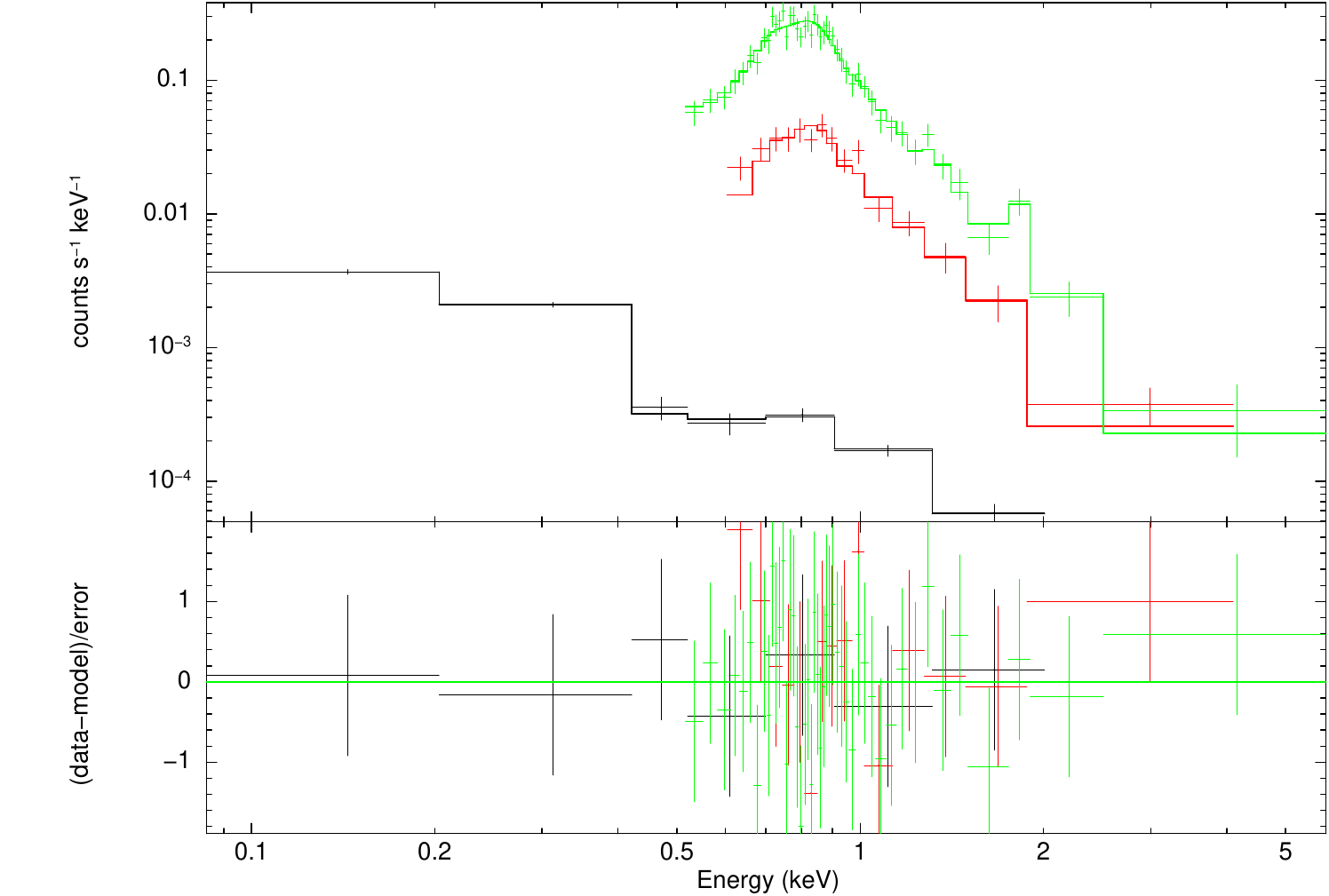}\label{fig:Group1SpecZ}}
\vspace{-0.3cm}\\
\subfloat[Group~2 within $0.2$--$0.5$ $R_{500}$]{\includegraphics[width=9cm]{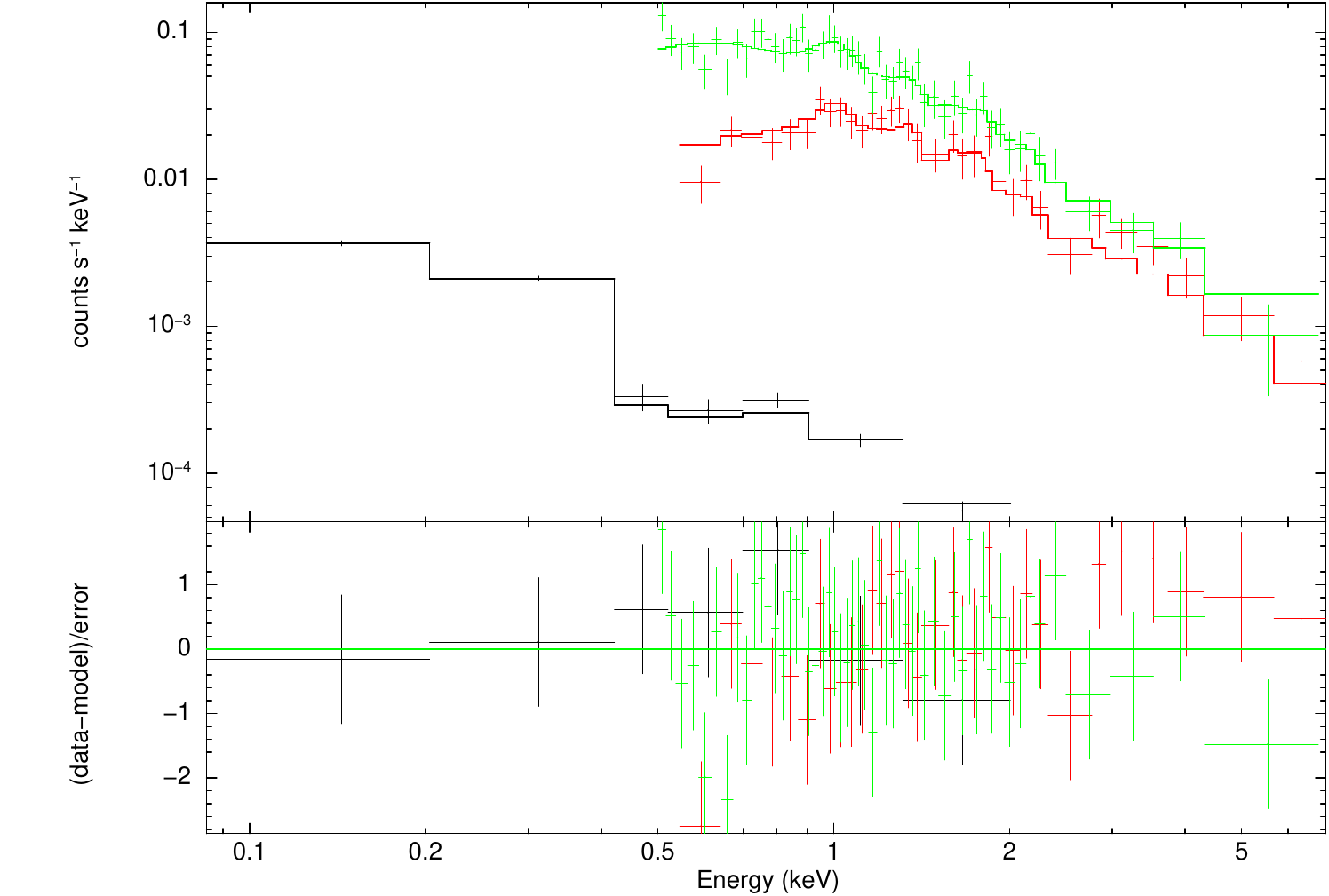}\label{fig:Group2Spec}}
\subfloat[Group~2 for $z$ estimation]{\includegraphics[width=9cm]{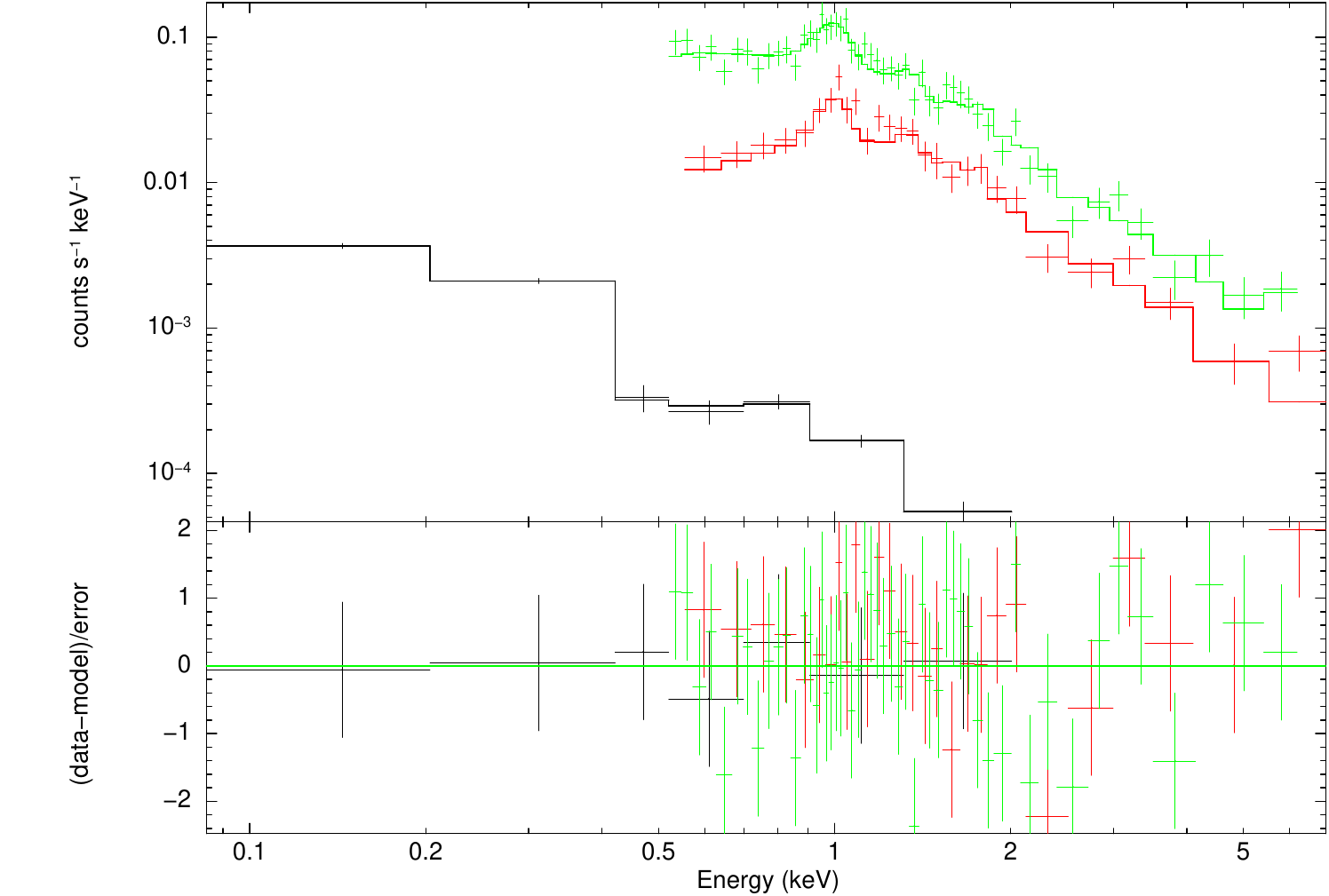}\label{fig:Group2SpecZ}}
\vspace{-0.3cm}\\
\subfloat[AS0924 within $0.2$--$0.5$ $R_{500}$]{\includegraphics[width=9cm]{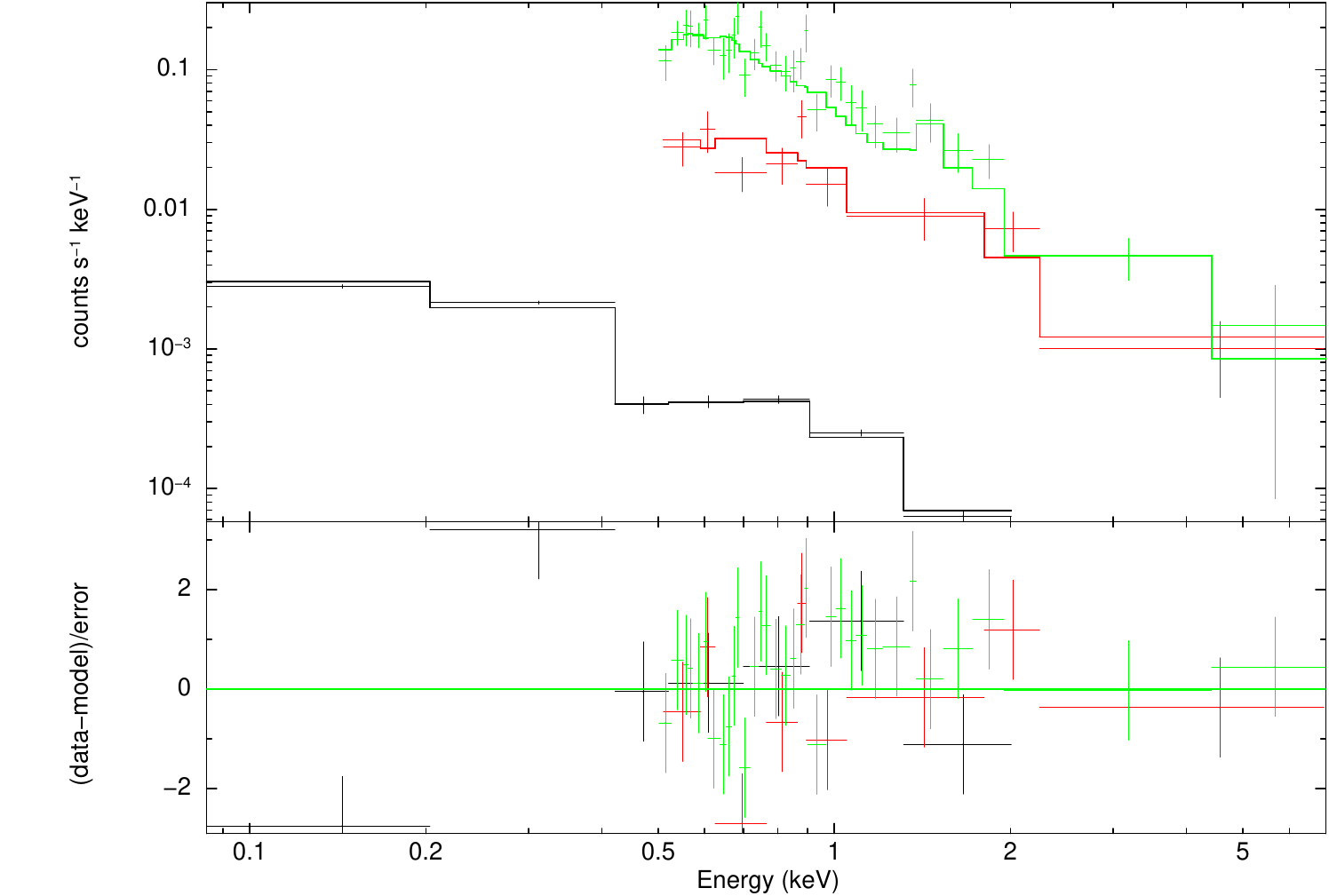}\label{fig:AS0924Spec}}
\subfloat[AS0924 for $z$ estimation]{\includegraphics[width=9cm]{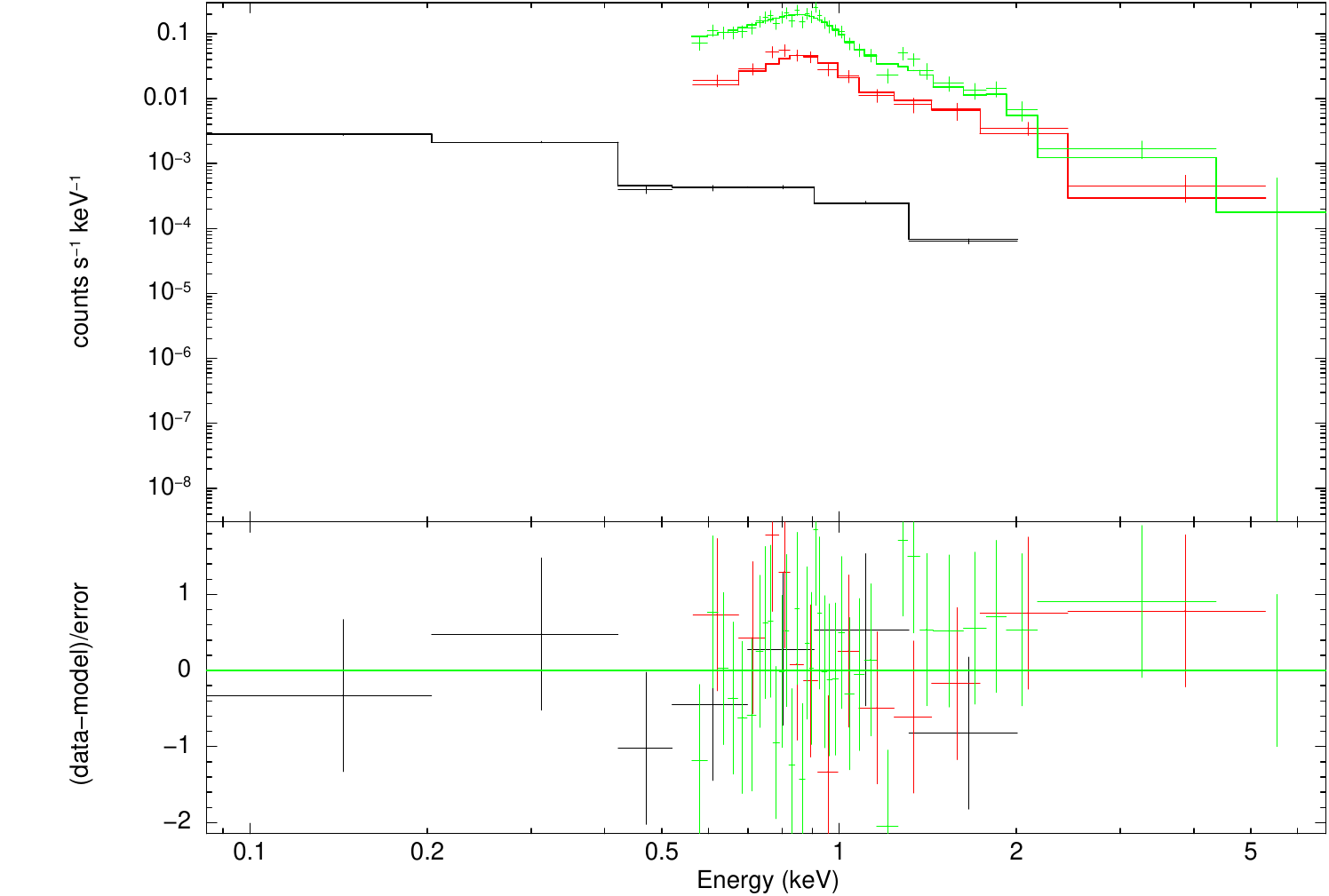}\label{fig:AS0924SpecZ}}
\\
\caption{EPIC-pn and -MOS2 spectra of the systems along with the best-fit models and residuals - part~1.}
\label{fig:Spectra1}
\end{figure*} 

\begin{figure*}[ht]
\centering
\subfloat[Group~3 within 1.6' (used for all fits)]{\includegraphics[width=9cm]{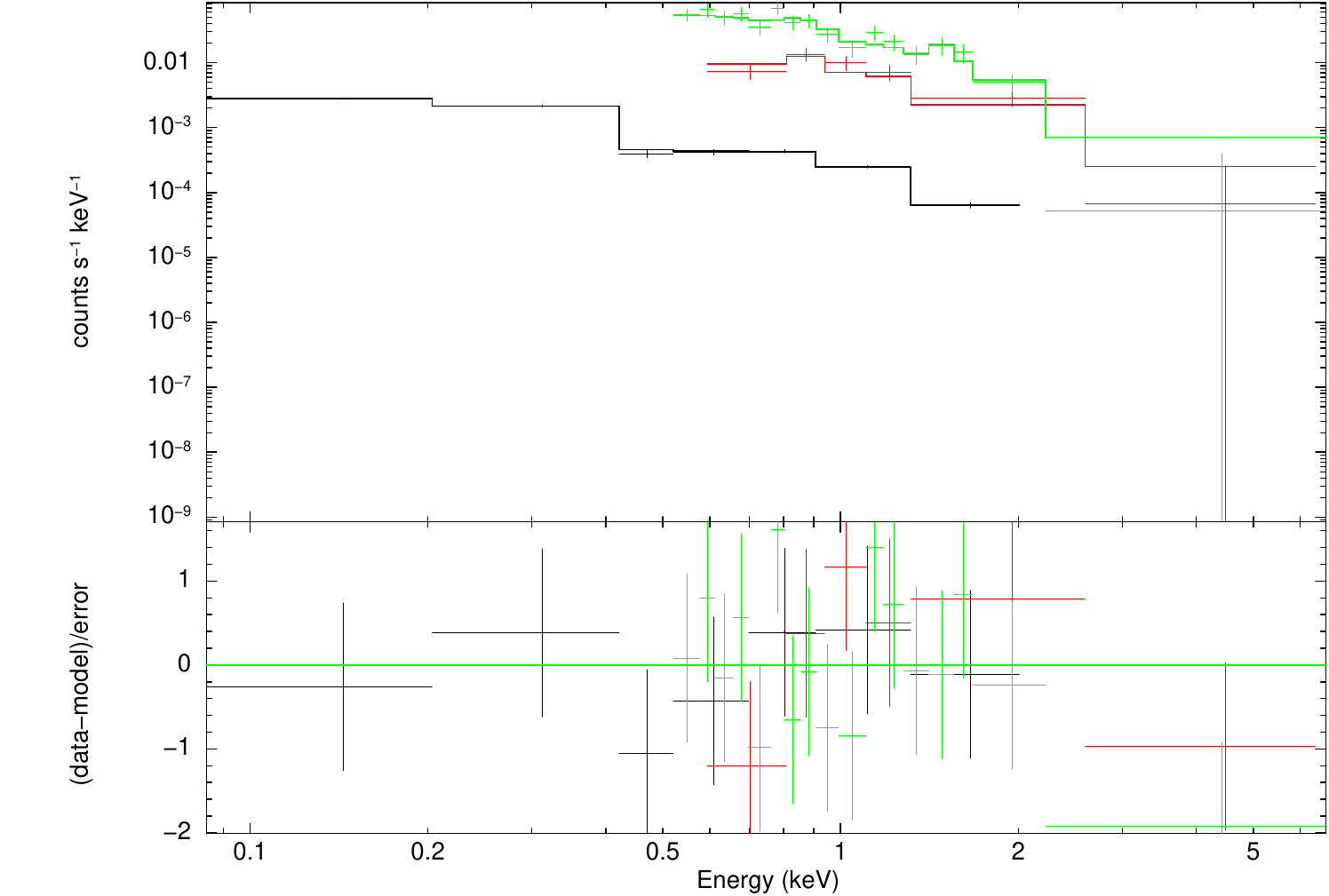}\label{fig:Group3Spec}}
\subfloat[A0349 for $z$ estimation]{\includegraphics[width=9cm]{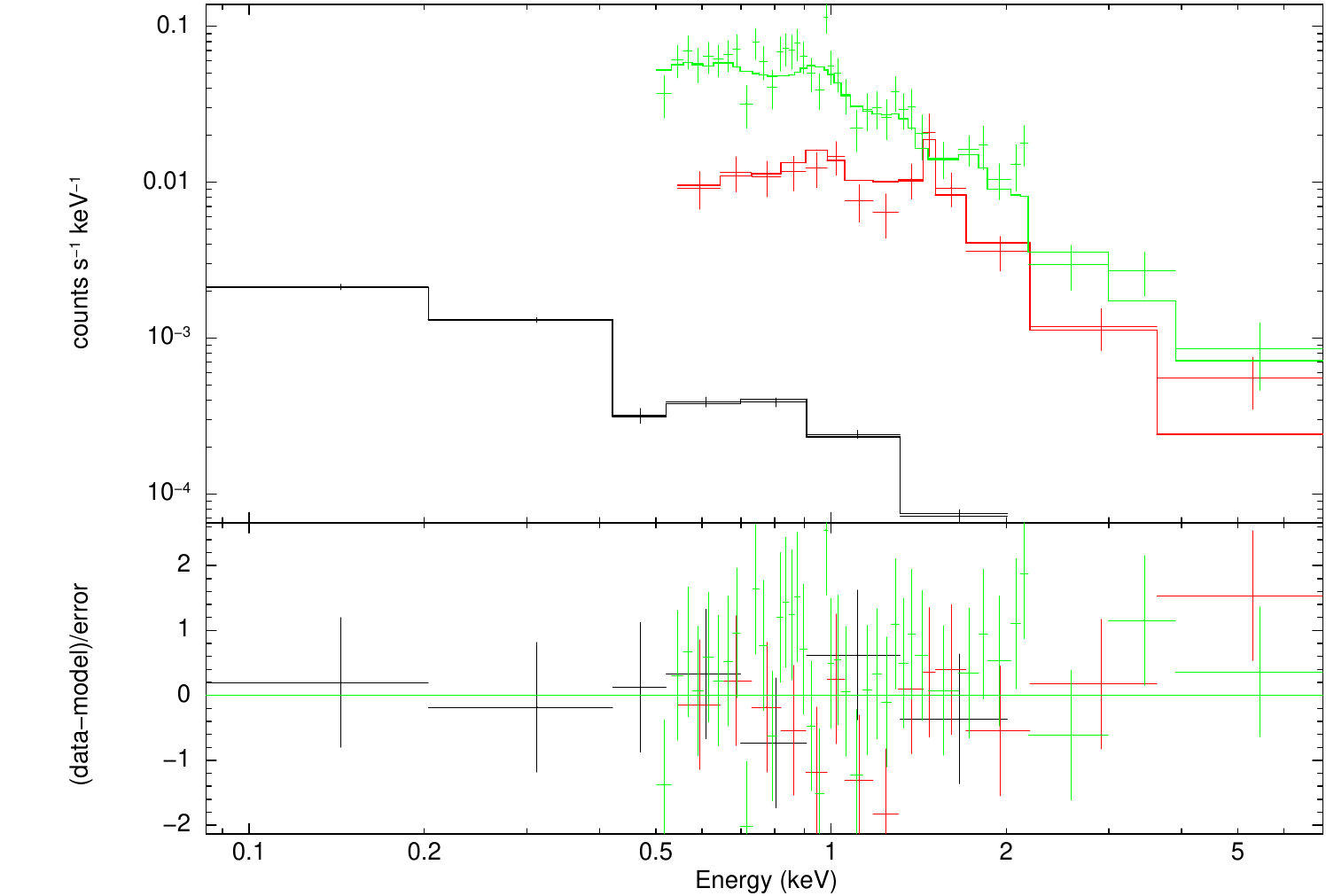}\label{fig:A0349SpecZ}}
\vspace{-0.3cm}\\
\subfloat[Group~4 within $0.2$--$0.5$ $R_{500}$]{\includegraphics[width=9cm]{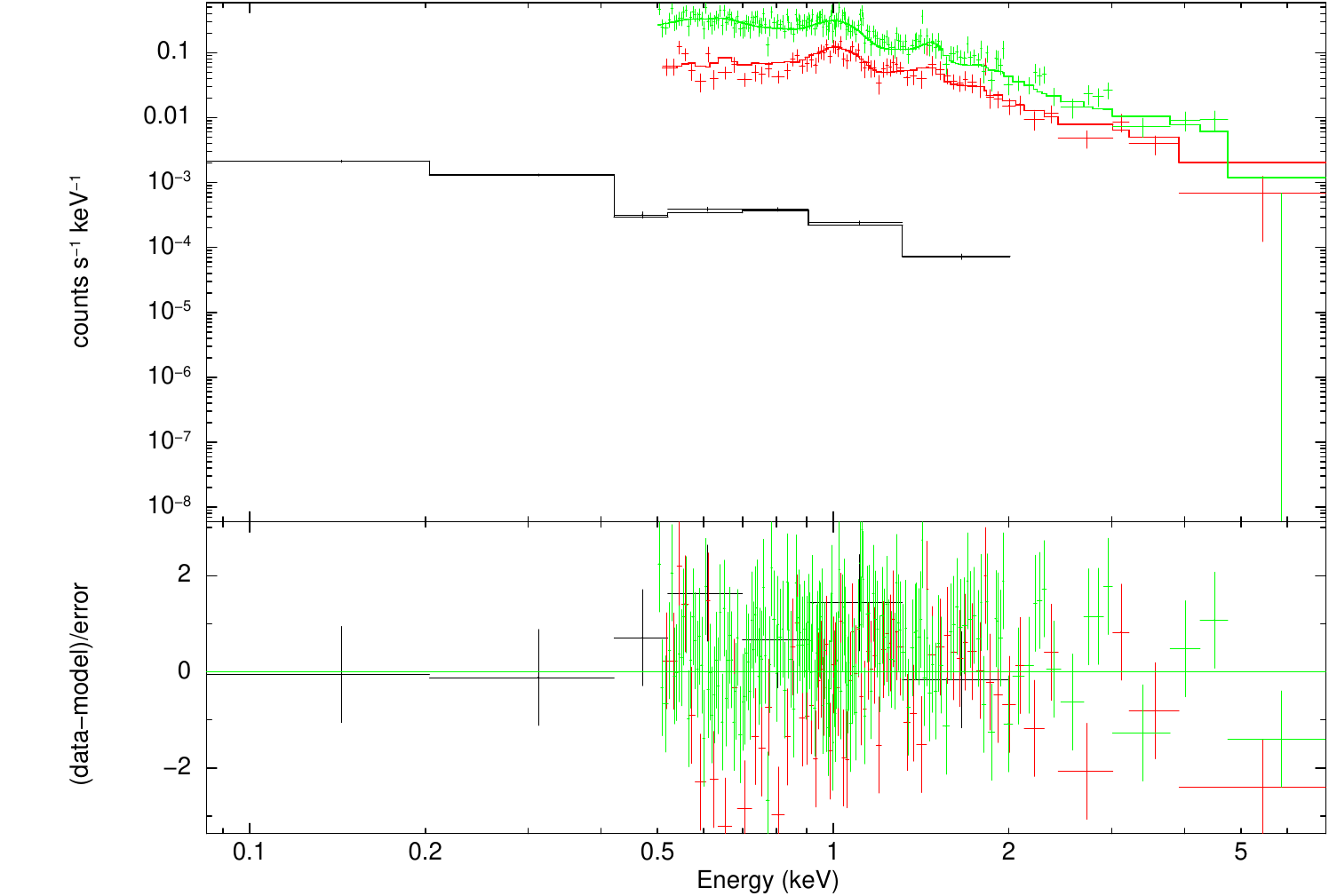}\label{fig:Group4Spec}}
\subfloat[Group~4 for $z$ estimation]{\includegraphics[width=9cm]{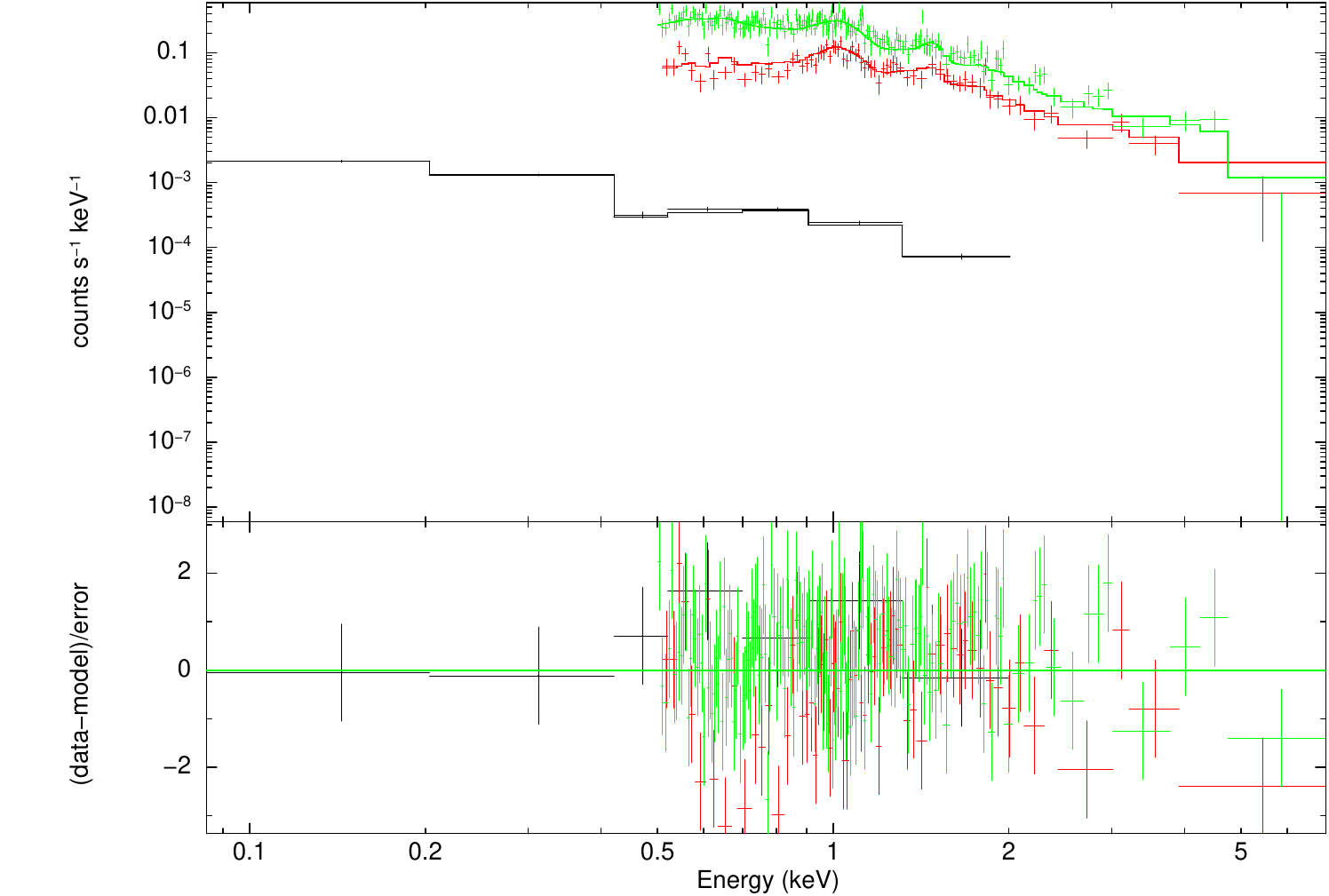}\label{fig:Group4SpecZ}}
\\
\caption{EPIC-pn and -MOS2 spectra of the systems along with the best-fit models and residuals - part~2.}
\label{fig:Spectra2}
\end{figure*} 

\end{document}